\input{psfig}
\documentclass[aps,final,notitlepage,oneside,twocolumn,nobibnotes,nofootinbib,
superscriptaddress,%
noshowpacs,%
centertags
]%
{revtex4i}
\newcommand{\be}{\begin{equation}}
\newcommand{\ee}{\end{equation}}
\newcommand{\bea}{\begin{eqnarray}}
\newcommand{\eea}{\end{eqnarray}}

\begin{document}
%
\title{Dark matter, density perturbations and structure formation}
\author{\firstname{A.} \surname{Del Popolo}}
\affiliation{Bo$\breve{g}azi$\c{c}i University, Physics Department,
80815 Bebek, Istanbul, Turkey }
\affiliation{Dipartimento di Matematica, Universit\`{a} Statale di Bergamo,
  via dei Caniana, 2,  24127, Bergamo, ITALY}
\affiliation{Istanbul Technical University, Ayazaga Campus,  Faculty of Science and Letters,  34469 Maslak/ISTANBUL, Turkey}

\noaffiliation
\begin{abstract}
\noindent {\bf Abstract}----This paper provides a review of the variants of dark matter which are thought to be fundamental components of the universe and their role in origin and evolution of structures and some new original results concerning improvements to the spherical collapse model. In particular, I show how the spherical collapse model is modified when we take into account
dynamical friction and tidal torques. 
\end{abstract}
\maketitle
\bigskip


\normalsize
\vspace{2cm}
\maketitle
%

\begin{flushleft}
{1. INTRODUCTION}
\end{flushleft}

The origin and evolution of large scale structure is today
the outstanding problem in cosmology. This is the most fundamental question
we can ask about the universe whose solution should help us to better understand problems
as the epoch of galaxy formation, the clustering in the galaxy
distribution, the amplitude and form of anisotropies in the microwave
background radiation. Several has been the approaches and models
trying to attack and solve this problem: no one has given a final answer. \\
The leading idea of all structure formation theories is that structures
was born from small perturbations in the otherwise uniform distribution
of matter in the early Universe, which is supposed to be, in great part,
dark (matter not detectable through light emission).

With the term Dark Matter cosmologists indicate an hypothetic material component of the universe which does not emit directly electromagnetic radiation (unless it decays in particles having this property (\cite{sci}, but also see \cite{stuart})).\\
Dark matter, cannot be revealed directly, but nevertheless it is necessary to postulate its existence in order to explain the discrepancies between the observed dynamical properties of galaxies and clusters of galaxies  
and the theoretical predictions based upon models of these objects assuming that the only matter present is the visible one.
If in the space were present a diffused material component having gravitational mass, but unable to emit electromagnetic radiation in significative quantity, this discrepancy could be eliminated (\cite{tur}). 
The study of Dark Matter has as its finality the explanation of formation of galaxies and in general of cosmic structures. For this reason, in the last decades, the origin of cosmic structures has been ``framed" in models in which Dark Matter constitutes the skeleton of cosmic structures and supply the most part of the mass of which the same is made.\\
There are essentially two ways in which matter in the universe can be revealed: by means of radiation, by itself emitted, or by means of its gravitational interaction with baryonic matter which gives rise to cosmic structures. 
Electromagnetic radiation permits to reveal only baryonic matter. In the second case, we can only tell that we are in presence of matter that interacts by means of gravitation with the luminous mass in the universe. 
The original hypotheses on Dark Matter go back to measures performed by Oort (\cite{oor}) of the surface density of matter in the galactic disk, which was obtained through the study of the stars motion in direction orthogonal to the galactic plane. The result obtained by Oort, which was after him named ``Oort Limit", gave a value of $ \rho = 0.15 M_{0} pc^{-3}$ for the mass density, and a mass, in the region studied, superior to that present in stars.
Nowadays, we know that the quoted discrepancy is due to the presence of HI in the solar neighborhood. Other studies (\cite{zwi}; \cite{smi}) showed the existence of a noteworthy discrepancy between the virial mass of clusters (e.g. Coma Cluster) and the total mass contained in galaxies of the same clusters.
These and other researches from the thirties to now, have confirmed that a great part of the mass in the universe does not emit radiation that can be directly observed. \\

\begin{flushleft}
{\it 1.1 Determination of $ \Omega $ and Dark Matter}
\end{flushleft}

The simplest cosmological model that describes, in a sufficient coherent manner, the evolution of the universe, from $ 10^{-2}
s$ after the initial singularity to now, is the so called {\it Standard Cosmological Model} (or Hot Big Bang model). It is based upon the Friedmann-Robertson-Walker (FRW) metric, which is given by:
\begin{equation}
ds^{2} = c^{2} d t^{2} -a(t)^{2}\left[\frac{d r^{2} }{1 - k r^{2}}+
r^{2} (d \theta^{2} +sin{\theta}^{2} d \phi^{2} )\right]
\end{equation}
where c is the light velocity, a(t) a function of time, or a scale factor called ``expansion parameter", t is the time coordinate, r, $\theta $ and  $ \phi $ the comoving space coordinates. The evolution of the universe is described by the parameter a(t) and it is fundamentally connected to the value $\rho$ of the average density.\\
The equations that describe the dynamics of the universe are the Friedmann's equations (\cite{fri}) that we are going to introduce in a while. These equations can be obtained starting from the equations of the gravitational field of Einstein (\cite{ein}): 
\begin{equation}
R_{ik}-\frac{1}{2} g_{ik} R =-\frac{8 \pi G}{ c^{4}} T_{ik}
\end{equation}
where now, $ R_{ik} $ is a symmetric tensor, also known as Ricci tensor, which describes the geometric properties of space-time, $ g_{ik} $ is the metric tensor, R is the scalar curvature, $T_{ik}$ is the energy-momentum tensor.\\
These equations connect the properties of space-time to the mass-energy. In other terms they describe how space-time is modeled by mass. Combining Einstein equations to the FRW metric leads to the dynamic equations for the expansion parameter, a(t). These last are the Friedmann equations:
\begin{equation}
d( \rho a^{3} )= -p d( a^{3} )
\end{equation}
\begin{equation}
\frac{1}{a^{2}} \dot{a}^{2} +\frac{k}{a^{2}} = \frac{8 \pi G }{3} \rho
\end{equation}
\begin{equation}
2\frac{\ddot{a}}{a} + \frac{\dot{a}^{2}}{a^{2}} +\frac{k}{a^{2}} =
-8 \pi G \rho
\end{equation}
where p is the pressure of the fluid of which the universe is constituted, k is the curvature parameter and a(t) is the scale factor connecting proper distances $ { \bf r }  $  to the comoving ones $ {\bf x} $ through the relation 
$ {\bf r}=a(t) {\bf x} $. One of the components of the today universe are galaxies. If we assume that galaxies motion satisfy  Weyl (\cite{wey}) postulate, the velocity vector of a galaxy is given by 
$ u^{i} = (1,0,0,0) $, and then the system behaves as a system made of dust for which we have $p=0$.
Only two of the three Friedmann equations are independent, because the first connects density, $\rho$ to the expansion parameter a(t). The character of the solutions of these equations depends on the value of the curvature parameter, $k$, which is also determined by the initial conditions by means of Eq. 3. 
The solution to the equations now written shows that if $\rho$ is larger than $ \rho_{c} = \frac{3 H^{2} }{ 8 \pi G } = 1.88* 10^{-29} g/cm^{3} $ (critical density, which can be obtained from Friedmann equations putting $t=t_0$, $k=0$, and $ H= 100 km /s Mpc $), space-time  has a closed structure ($k=1$) and equations shows that the system go through a singularity in a finite time. This means that the universe has an expansion phase until it reaches a maximum expansion after which it recollapse. If $ \rho < \rho_{c} $, the expansion never stops and the universe is open $k=-1$ (the universe has a structure similar to that of an hyperboloid, in the two-dimensional case).
If finally, $ \rho = \rho_{c} $ the expansion is decelerated and has infinite duration in time, $k=0$, and the universe is flat (as a plane in the two-dimensional case). The concept discussed can be expressed using the parameter $ \Omega = \frac{\rho}{\rho_{c}} $. In this case, the condition $ \Omega = 1 $ corresponds to $k=0$, $ \Omega <1 $ corresponds to $k=-1$, and $ \Omega > 1 $ corresponds to $k=1$. \footnote{See next paragraphs for some items on cosmological models with non-zero cosmological constant.}\\
The value of $ \Omega $ can be calculated in several ways. The most common methods are the dynamical methods, in which the effects of gravity are used, and kinematics methods sensible to the evolution of the scale factor and to the space-time geometry. The results obtained for $\Omega$ with these different methods are summarized in the following.\\

{\it Dynamical methods}:\\

(a) Rotation curves: The contribution of spiral galaxies to the density in the universe is calculated by using their rotation curves and the third Kepler law. Using the last it is possible to obtain the mass of a spiral galaxy from the equation:
\begin{equation}
M(r) = v^{2} r /G  \label{eq:kep}
\end{equation}
where v is the velocity of a test particle at a distance r from the center and M(r) is the mass internal to the circular orbit of the particle. In order to determine the mass M is necessary to have knowledge of the term  $v^{2} $ in Eq. (\ref{eq:kep}) and this can be done from the study of the rotation curves through the 21 cm line of HI.  
Rotation curves of galaxies are characterized by a peak reached at distances of some Kpcs and a behavior typically flat for the regions at distance larger than that of the peak. A peculiarity is that the expected Keplerian fall is not observed. This result 
is consistent with extended haloes containing masses till 10 times the galactic mass observed in the optical (\cite{van}). The previous result is obtained 
assuming that the halo mass obtained with this method is distributed in a spherical region so that we can use Eq. (\ref{eq:kep}) and that we neglect the tidal interaction with the neighboring galaxies which tend to produce an expansion of the halo.
After M and the luminosity of a series of elliptical galaxies is determined, the contribution to the density of the universe is given by $ \rho = < \frac{M}{L} > \ell $ where $\ell$ is the luminosity per unit volume due to galaxies 
and can be obtained from the galactic luminosity function $ \phi(L) dL,  $ which describes the number of galaxies per $ Mpc^{3} $ and luminosity range $ L, L+dL $. The value that is usually assumed for $\ell$ is $ \ell= 2.4h 10^{8} L_{bo} Mpc^{-3} $. The arguments used lead to a value of $\Omega_{g}$ for the luminous parts of spiral galaxies of $\Omega_{g} \leq 0.01$, while for haloes $ \Omega_{h} \geq 0.03 -0.1 $. The result shows that the halo mass is noteworthy larger than the galactic mass observable in the optical (\cite{pee}).\\

{(b) \it Virial theorem}:\\

In the case of non spiral galaxies and clusters, the mass can be obtained using the virial theorem $ 2 T + V = 0 $, with 
\begin{equation}
T \cong \frac{3}{2} M <v_{r}^{2}> 
\end{equation}
where $ < v_{r}^{2} > $ is the velocity dispersion along the line of sight. After getting the value of M of the cluster by means of the virial theorem one determines L by means of observations. Given M and L, the value of $ \Omega $ for clusters is obtained similarly to the case of spiral galaxies. Usual values obtained for $\Omega$ are $ \Omega = 0.1-0.3 $ (\cite{pee}). A problem of the quoted method is that in general the results obtained are the right one only for virialized, spherically symmetric clusters. In general, clusters are not virialized objects: even Coma clusters seems to have a central core constituted by more than one blob of mass (\cite{henry}).  \\

{(c) \it Peculiar velocities}: \\

The velocity of a galaxy can be written as:
\begin{equation}
{\bf V}_{g} =H{\bf r} + \delta{\bf v}
\end{equation}
where $H$ is Hubble constant. The previous equation shows that the motion of a galaxy is constituted by two components:
the velocity of the galaxy due to the Hubble flow and a peculiar velocity $\delta{\bf v}$, which describes the motion of the galaxy with respect to the background. In the linear regime, as we see in a while, we find that on average on a scale of length $\lambda$, it is:
\begin{equation}
\frac{\delta v}{c} \approx \Omega^{0.6}
\frac{\lambda}{H_{0}^{-1}}\frac{\delta \rho}{\rho}
\end{equation}
(\cite{kol}).
Then given the overdensity $ \frac{\delta \rho}{\rho} $ on scale $ \lambda $ and $ \delta v $, it is possible to obtain $\Omega$. The overdensity $ \frac{\delta \rho}{\rho} $ can be obtained from the overdensity of galaxies $ \frac{\delta n_{g}}{n_{g}}$ using the relation $ \frac{\delta \rho}{\rho} = \frac{\delta n_{g} }{n_{g}} b^{-1} $
with $ 1<b<3 $. Using IRAS catalog in order to obtain the overdensity in galaxies one finds $ \Omega \cong 1 $. The values of $\Omega$ obtained using the method of peculiar velocity assume that the peculiar velocity fields describe in an accurate way, the inhomogeneity in the distribution of underlying mass. 
We should note that the peculiar velocity method has some difficulties. In general, in order to obtain these last it is necessary to determine the redshift and the distance of galaxies and by using these data it is possible to obtain the peculiar velocity:   
\begin{equation}
v_{pec} = z c - H_{0} d
\end{equation}
It is evident that there are problems in measuring the distance d, problems connected to difficulties in finding  trustable indicators of distance. Moreover the peculiar velocity can be determined only along the line of sight. \\

{(d) \it Kinematic methods}:\\

These methods are based upon the use of relations between physical quantities dependent on cosmological parameters. An example of those relations is the relation luminosity distance-redshift:
\begin{equation}
H_{0} d_{L} = z+\frac{1}{2}(1-q_{0} )z^{2}
\end{equation}
where $ H_{0} $ is Hubble constant nowadays, z is the redshift, $ d_{L} = \frac{4 \pi L}{F} $ the luminosity distance, L the absolute luminosity, and F the flux. 
By means of the relations luminosity-redshift, angle-redshift, number of objects-redshift, it is possible to determine the parameter of deceleration $ q_{0} = -\frac{\ddot{a_{0}}}{H_{0}^{2} a_{0}} $ ($ a_{0} $ and 
$ H_{0}=\frac{\dot{a_{0}}}{a_{0}}=100 h km /Mpc s $ are the scale factor and the Hubble constant, nowadays). 
At the same time $ q_{0} $ is connected to $ \Omega $ by means of $ q_{0} = \frac{\Omega}{2} $, in a matter universe.
One of the first test used, the luminosity distance-redshift has several problems due to effects of the evolution of sources. Uncertainties in the knowledge of the effects of galactic evolution on the intrinsic luminosity of the same has not, in the past, permitted to find definitive values of  $ q_{0} $. For this reason, \cite{loh} introduced another kinematic test: number of galaxies-redshift. This test is based on the count of the number of galaxies in a comoving element of galaxies, defined by the surface $ d \Omega $ and the redshift $d z$. 
This number depends on $ q_{0} $. Nevertheless the effects of evolution of sources influences on the results of the test, it is more sensible to the evolution of number of sources than to the evolution of luminosity, on which there is not an accepted theory. Results gives high values of $\Omega$ ($\Omega =0.9_{-0.5}^{+0.7} $) (\cite{kol}). 
\footnote{Among kinematics methods we should mention SNeIa which played a key role in the last few years (\cite{val}).}\\

{(e) \it Primordial nucleosynthesis}:\\
The theory of primordial nucleosynthesis, proposed in 1946 by Gamov, assumes that the light elements till $ Li_{7} $ are generated after big bang and that heavier elements originate from nuclear reactions inside stars. The values obtained for the abundances depends on some parameters like: $ \eta $, the value of the ratio baryons-photons, nowadays; $ N_{\nu} $, the number of neutrinos species; $ T_{CMBR} $, the temperature of Cosmic Microwave Background Radiation. 
The theory of primordial nucleosynthesis permits to give limits to $\Omega_b$ (b stands for baryons). With a ratio baryons-photons $ 3 *10^{-10} \leq \eta \leq 5 *10^{-10} $, and a value of $ N_{\nu} \leq 4 $ for the neutrinos species, $ T_{CMBR}= 2.736 \pm 0.01 K $, is found $ 0.011\leq \Omega_{b} \leq 0.12 $ (\cite{kol}).\\


In the following, we summarize the results of some more recent results. 

\cite{dek2} used several methods to obtain the value of $\Omega$.
According to their classification, we divide the methods into the following four classes:

\leftskip=1.0 true cm

\leftskip=1.0 true cm
\noindent
{\hskip -0.5 true cm}
$\bullet\ ${\it Global measures}. Based on properties of space-time
that constrain combinations of $\Omega_m$ and the
other cosmological parameters ($\Lambda$, $H_0$, $t_0$).

\leftskip=1.0 true cm
\noindent
{\hskip -0.5 true cm}
$\bullet\ ${\it Virialized Systems}. Methods based on nonlinear dynamics 
within galaxies and clusters on comoving scales $1-10 h^{-1} Mpc$.

\leftskip=1.0 true cm
\noindent
{\hskip -0.5 true cm}
$\bullet\ ${\it Large-scale structure}. Measurements based on 
mildly-nonlinear gravitational dynamics of fluctuations on scales 
$10-100 h^{-1} Mpc$ of superclusters and voids, in particular {\it cosmic flows}.

\leftskip=1.0 true cm
\noindent
{\hskip -0.50 true cm}
$\bullet\ ${\it Growth rate of fluctuations}. Comparisons of present day
structure with fluctuations at the last scattering of the cosmic microwave 
background (CMB) or with high redshift objects of the young universe.

\leftskip=0.0 true cm

The methods and current estimates are summarized in Table 3.
The estimates based on virialized objects typically yield low values of
$\Omega_m \sim 0.2-0.3$. The global measures, large-scale structure and cosmic
flows typically indicate higher values of $\Omega_m \sim 0.4-1$.

Bahcall et al. (\cite{bah}),
showed that the  evolution of the number density of rich clusters of galaxies
breaks the degeneracy between $\Omega$ (the mass density
ratio of the universe) and
$\sigma_{8}$ (the normalization of the power spectrum), $\sigma_{8} \:
\Omega^{0.5} \simeq 0.5$, that
follows from the observed present-day abundance of rich clusters.  The
evolution of high-mass
(Coma-like) clusters is strong in $\Omega = 1$, low-$\sigma_{8}$ models (such as
the standard biased CDM model with $\sigma_{8} \simeq 0.5$), where the
number density of clusters
decreases by a factor of $\sim 10^{3}$ from $z = 0$ to $z \simeq 0.5$; the
same clusters show only mild evolution in low-$\Omega$, high-$\sigma_{8}$ models,
where the decrease is
a factor of $\sim 10$.
This diagnostic provides a most powerful constraint on $\Omega$.
Using observations of clusters
to $z \simeq 0.5-1$, the authors found  
only mild evolution in the observed cluster abundance, and 
$\Omega = 0.3 \pm 0.1$ and $\sigma_{8} = 0.85 \pm 0.15$
(for $\Lambda = 0$ models;
for $\Omega + \Lambda = 1$ models, $\Omega = 0.34 \pm 0.13$). 

ferreira et al. (\cite{fer}), proposed an
alternative method to estimate $v_{12}$ directly from peculiar velocity
samples, which contain redshift-independent distances as well as galaxy
redshifts.
In contrast to other dynamical
measures which determine $\beta\equiv\Omega^{0.6}\sigma_8$, this method
can provide an estimate of $\Omega^{0.6}\sigma_8^2$
for a range of $\sigma_8$
where $\Omega$ is the cosmological density parameter, while $\sigma_8$
is the standard normalization for the power spectrum of density
fluctuations. 

Melchiorri (\cite{mel}), used the angular power spectrum of the Cosmic Microwave Background,
measured during the North American test flight of the BOOMERANG experiment,
to constrain the geometry of the universe. Within the class
of Cold Dark Matter models, they find 
that the overall fractional energy density of the universe, $\Omega$, is
constrained to be $0.85 \le \Omega \le 1.25$ at the $68\%$ confidence level.

Branchini (\cite{branchi}), compared 
the density and velocity fields as extracted from the Abell/ACO
clusters to the corresponding fields recovered by 
the POTENT method from the Mark~III peculiar velocities of galaxies.
Quantitative comparisons within a volume containing $\sim\!12$ 
independent samples yield $\beta_c \equiv \Omega^{0.6}/b_c=0.22\pm0.08$, 
where $b_c$ is the cluster biasing parameter at $15 h^{-1} Mpc$. If $b_c
\sim 4.5$, as indicated by the cluster correlation function, their
result is consistent with $\Omega \sim 1$.

{(f) \it Inflation}:\\

It is widely supposed that the very early universe experienced an era of
inflation (see \cite{guth}, \cite{lin}, \cite{kol}). By `inflation' one means 
that the scale factor has positive
acceleration, $\ddot a>0$, corresponding to repulsive gravity and $3p<-\rho$.
During inflation $aH=\dot a$ is increasing, so that comoving scales are
leaving the horizon (Hubble distance) rather than entering it, and it is
supposed that at the beginning of inflation the observable universe was well
within the horizon.

The inflationary hypothesis is attractive because it holds out the possibility
of calculating cosmological quantities, given the Lagrangian describing the
fundamental interactions. The Standard Model, describing the interactions up
to energies of order $1 TeV$, is not viable in this context because it does
not permit inflation, but this should not be regarded as a serious setback
because it is universally agreed that the Standard Model will require
modification at higher energy scales, for reasons that have nothing to do with
cosmology. The nature of the required extension is not yet known, though it is
conceivable that it could become known in the foreseeable future. 
But even without a specific model of the interactions (ie., a specific
Lagrangian), the inflationary hypothesis can still offer guidance about what
to expect in cosmology. More dramatically, one can turn around the
theory-to-observation sequence, to rule out otherwise reasonable models.
The importance of inflation is connected to:\\
a) the origin of density perturbations, which 
could originate during inflation as quantum fluctuations, which become
classical as they leave the horizon and remain so on re-entry. The original
quantum fluctuations are of exactly the same type as those of the
electromagnetic field, which give rise to the experimentally observed Casimir
effect.\\
b) One of the most dramatic and
simple effects is that there is no fine-tuning of the initial value of the
density parameter $\Omega=8\pi\rho/3 m_{Pl}^2 H^2$. From the
Friedmann equation, $\Omega$ is given by
\begin{equation} 
\Omega-1=(\frac{K}{aH})^2 \label{51} 
\end{equation}
Its present value $\Omega_0$ is certainly within an order of magnitude of 1,
and in the absence of an inflationary era $\Omega$ becomes ever smaller as one
goes back in time, implying an initial fine tuning. In contrast, if there is
an inflationary era beginning when the observable universe is within the
horizon, Eq. (\ref{51}) implies that $\Omega_0$ will be of order 1, provided only
that the same is true of $\Omega$ at the beginning of inflation. A value of
$\Omega_0$ extremely close to 1 is the most natural, though it is not
mandatory.\footnote {An
argument has been given for $\Omega_0$ very close to 1 on the basis of effects
on the cmb anisotropy from regions far outside the observable universe (\cite{tur}), but it is not valid as it stands because it ignores spatial curvature.}\\
c) Another effect of inflation is that it can eliminate particles and topological
defects which would otherwise be present. Anything produced before inflation
is diluted away, and after inflation there is a maximum temperature (the
`reheat' temperature) which is not high enough to produce all the particles
and defects that might otherwise be present. As we shall remark later, this
mechanism can remove desirable, as well as undesirable, objects.\\
d) 
The most dramatic effect of inflation is that it may offer a way of
understanding the homogeneity and isotropy of the universe, or at any rate of
significant regions of it. We have nothing to say about this complex issue in
its full generality,  but a more modest version of it is our central concern.
In this version, one begins the discussion at some early stage of inflation,
when the universe is supposed already to be {\it approximately} homogeneous
and isotropic.
 One then argues that in that case, scales far inside the
horizon must be {\it absolutely} homogeneous and isotropic, except for the
effect of vacuum fluctuations in the fields. Finally, one shows that after
they leave the horizon, such fluctuations can become the classical
perturbations that one deals with in cosmological perturbation theory. This
possibility was first pointed out for gravitational waves by 
\cite{staa} and for density perturbations by several people (\cite{guth1};
\cite{haw}; \cite{sta}). As we shall go to some trouble to demonstrate,
the vacuum fluctuations can be evaluated unambiguously once an inflationary
model is specified.

{(g) \it Scalar field inflation}:\\

Two mechanisms for inflation have been proposed. The simplest one
(\cite{guth}) invokes a scalar field, termed the inflaton field.
An alternative (\cite{staa}) is to invoke a modification of
Einstein gravity, and
combinations of the two mechanisms have also been proposed. During inflation
however, the proposed modifications of gravity
can be abolished by redefining the spacetime metric tensor, so that one
recovers the scalar field case. We focus on it for the moment,
but modified gravity models
will be included later in our survey of
specific models.

In comoving coordinates a homogeneous scalar field
 $\phi$ with minimal coupling to gravity has the
equation of motion
\begin{equation} 
\ddot \phi+3 H\dot \phi +V^\prime (\phi) =0 \label{52} 
\end{equation}
Its energy density and pressure are 
\begin{eqnarray}
\rho&=& V+\frac12\dot\phi^2\\
p&=&-V +\frac12\dot\phi^2 
\end{eqnarray}
If such a field dominates $\rho$ and $p$, the inflationary condition
$3p<\rho$ is achieved provided that the field rolls sufficiently slowly,\begin{equation} \dot\phi^2<V \end{equation}

Practically all of the usually considered models of inflation satisfy three
conditions. First, the motion of the field is overdamped, so that the `force'
$V^\prime $ balances the `friction term' $3H\dot\phi$,
\begin{equation}
\dot{\phi} \simeq -\frac{1}{3H} V' \label{56}
\end{equation}
Second,
\begin{equation}
\epsilon \equiv \frac{m_{Pl}^2}{16\pi}
\left( \frac{V'}{V} \right)^2 \ll 1 \label{57} 
\end{equation}
which means that the inflationary requirement $\dot\phi^2<V$ is well
satisfied and
\begin{equation} H^2 \simeq \frac13 \frac{8\pi}{m_{pl}^2} V \label{57a} \end{equation}
These two conditions imply that
$H$ is slowly varying, and that the scale factor increases more or less
exponentially,
\begin{equation} a\propto e^{Ht} \label{58b} \end{equation}
The third condition that is usually satisfied is
\begin{equation} |\eta|\ll1 \label{58} 
\end{equation}
where
\begin{equation} 
\eta \equiv \frac{m_{Pl}^2}{8\pi} \frac{V''}{V} 
\end{equation}
It can be `derived' from the other two by differentiating
the approximation Eq. (\ref{56}) for $\dot\phi$ and noting that consistency with the
exact expression  Eq. (\ref{52}) requires $\ddot \phi\ll V^\prime $ is satisfied.
However
there is no logical necessity for the derivative of an approximation to be
itself a valid approximation, so this third condition is logically independent
of the others. Conditions involving higher derivatives of $V$ could be
`derived' by further differentiation, with the same caveat, but the two that
we have given, involving only the first and second derivatives, are the ones
needed to obtain the usual predictions about inflationary perturbations. The
term `slow-roll inflation' is generally taken to denote a model in which they
are satisfied and we are adopting that nomenclature here. Practically all of
the usually considered models of inflation satisfy the slow-roll conditions
more or less well.

It should be noted that the first slow-roll condition is on a quite different
footing from the other two, being a statement about the {\em solution} of the
field equation as opposed to a statement about the potential that
defines this equation. What we are saying is that in the usually
considered models one can show
that the first condition is an attractor solution, in a regime typically
characterized by the other two conditions, and that moreover reasonable initial
conditions on $\phi$ will ensure that this solution is achieved well before
the observable universe leaves the horizon.
It is important to remember that
there are strong observational limits for the parameters previously introduced (e.g. $\epsilon$, $\eta$).
For example \cite{mel1}
studied the possible contribution of a stochastic gravitational
wave background to the anisotropy of the cosmic microwave background in cold and mixed dark matter (CDM and MDM) models.
This contribution was tested against detections of CMB anisotropy
at large and intermediate angular scales. 
The best fit parameters (i.e. those which
maximize the likelihood) are (with $ 95 \%$ confidence)
$n_S=1.23^{+0.17}_{-0.15}$
and
\begin{equation}
 R(n_S)=\frac{C_2^T}{C_2^S} = {{29 \epsilon} \over{ \pi^2 f(n_S)}}=2.4^{+3.4}_{-2.2}
\end{equation}
where
\begin{equation}
f(n_S)=\frac{\displaystyle \Gamma(3-n_S)\Gamma({3+n_S \over
2})}
{\displaystyle\Gamma^2({4-n_S  \over 2})\Gamma({9-n_S \over
2})}
\end{equation}
The previous constraint fixes the value of $\epsilon$ as well that of $\eta$ 
\begin{equation}
2 \eta=n_s-1+2 \epsilon
\end{equation}

They find that by including the possibility of such background in CMB data
analysis it can drastically alter the conclusion on the remaining
cosmological parameters.
More stringent constraints on some of the previous parameters are given in section 1.12.


{(h) \it Conclusions}:\\
We have seen the possible values of $\Omega$ using different methods. We have to add that Cosmologists are ``attracted" 
by a value of $\Omega_0=1$. This value of $\Omega$ is requested by inflationary theory.
The previous data lead us to the following hypotheses:\\
i) $ \Omega_{0} < 0.12 $; in this case one can suppose that the universe is fundamentally made of baryonic matter (black holes; Jupiters; white dwarfs).\\
ii) $\Omega_{0} > 0.12$; in this case in order to have a flat universe, it is necessary a non-baryonic component. 
$\Omega_b=1$ is excluded by several reasons (see \cite{efs}, \cite{kol}.
The remaining possibilities are:\\
1) existence of a smooth component with $ \Omega = 0.8 $. \\
The test of a smooth component can be done with kinematic methods. \\ 
2) Existence of a cosmological term, absolutely smooth to whom correspond an energy density
$ \rho_{vac} =\frac{ \Lambda }{8 \pi G } $. \\
3) existence of non-baryonic matter: the universe is fundamentally done of particles (neutrinos, WIMPS (Weakly Interacting Massive Particles)). \\
4) A combination of 2) and 3).\\
Before going one, I want to recall that some authors (\cite{fin}; \cite{mil}; \cite{san}) have assumed that we have a scant knowledge of physical laws. Sanders assumes that the gravitational potential changes with distance and in particular the gravitational constant has a different value at large distances. Milgrom assumes that the Newton law of gravitation is not valid when the gradient of the potential is small. In this case, the problem of the dynamics of clusters of galaxies is solved without introducing Dark Matter. In any case, the quoted assumptions have no general theory that can justify them.

\begin{flushleft}
{\it 1.2 Dark matter in particles}
\end{flushleft}


We know that if $ \Omega=1 $, dark matter cannot be constituted exclusive of baryonic matter. The most widespread hypothesis is that dark matter is in form of particles. Several candidates exist: neutrinos, axions, neutralinos, photinos, gravitinos, etc. Interesting particles are usually grouped into three families:\\
HDM (Hot Dark Matter), CDM (Cold Dark Matter) and WDM (Warm Dark Matter). In order to understand this classification it is necessary to go back to the early phases of universe evolution. The history of the universe is characterized by long phases of local thermodynamic equilibrium (LTE) and by ``deviation" by it: nucleosynthesys, bariogenesys, decoupling of species, etc. In the early universe, were present the particles that we know today and other particles predicted theoretically, but that have not been observed. Massive particles preserved the thermodynamic equilibrium concentration 
until the rate, $ \Gamma $, of reactions and interactions that produced that concentration was larger than the expansion rate of the universe, H. When the condition $ \Gamma > H $ was no longer satisfied the reactions stopped and the abundance of the considered species remained constant at the value it had at time of freez-out, $ T_{f} $, time at which $ \Gamma = H $. If we indicate with $ Y=\frac{n}{s} $ the number of particles per unit comoving volume and we remember that n is the number density of species and s the entropy density, we obtain a contribution of the species to the actual density of the universe as $ \Omega h^{2} =0.28 Y(T_{f} )(\frac{m}{ev}) $ (\cite{kol}). At time $ T_{f} $, particles could be relativistic or non-relativistic. Relativistic particles are today indicated with the term hot cosmic relics, HDM, while non-relativistic particles are named cold cosmic relics, CDM. There is an intermediate case, that of warm relics, WDM.\\
An example of HDM are massive neutrinos. Possible masses for these neutrinos are: 
\begin{equation}
25 ev \leq m_{\nu} \leq 100 ev 
\end{equation}
(\cite{sci}) and 
\begin{equation}
m_{\nu} \geq \left\{\begin{array}{ll}
4.9-1.3 Gev & \mbox{for Maiorana's neutrinos}\\
1.3-4.2 Gev & \mbox{for Dirac's neutrinos}
\end{array}\right.
\end{equation}
(\cite{lee}).\\
There are confirmed experimental evidence of the existence of massive neutrinos. In 1980, \cite{lyu} announced the detection of an electronic antineutrino with mass 30 eV, by means of the shape of the electron energy spectrum in the $\beta$ decay of tritium (\cite{jel}).
Experiments (Super-Kamiokande, SNOW) have obtained some evidence 
of non-zero mass from neutrino oscillations. This yields a difference of square 
masses of order $10^{-3}$ eV, and a mass of 0.05 eV (in the simplest case) (see \cite{zu}). \\
Among typical examples of CDM we have WIMPS and in particular axions and neutralinos (SUSY particle). This particle was postulated in order to solve the strong CP problem in nuclear physics. This problem arises from the fact that some interactions violate the parity, P, time inversion, T, and CP. If these are not eliminated, they give rise to a dipole momentum for the neutron which is in excess of ten order of magnitude with respect to experimental limits (\cite{kol}). The solution to the problem was proposed by  Peccei-Quinn in 1977 (\cite{pec}) in terms of a spontaneous symmetry breaking scheme. To this symmetry breaking should be associated 
a Nambu-Goldstone boson: the axion. The axion mass ranges between $ 10^{-12} ev $-$ 1 Gev $. In cosmology there are two ranges of interest: $ 10^{-6} ev \leq m_{a} \leq 10^{-3}ev $ ; $ 3 ev \leq m_{a} \leq 8 ev $.
Axion production in the quoted range can originate due to a series of astrophysical processes (\cite{kol}) and several are the ways these particles can be detected. Nevertheless the effort of researchers expecially in USA, Japan and Italy, axions remain hypothetical particles. They are in any case the most important CDM candidates.\\

In the following, I am going to speak about the basic ideas of structure formation. I shall write about density perturbations, their spectrum and evolution, about correlation functions and their time evolution, etc.

\begin{flushleft}
{\it 1.3 Origin of structures}
\end{flushleft}


Observing our universe, we notice a clear evidence of inhomogeneity when we consider small scales (Mpcs). In clusters density reaches values of $10^3$ times larger than the average density, and in galaxies it has values $10^5$ larger than the average density (\cite{kol}). If we consider scales larger than $10^2$ Mpcs universe appears isoptric as it is observed in the radio-galaxies counts, in CMBR, in the X background (\cite{pee}). The isotropy at the decoupling time, $t_{dec}$, at which matter and radiation decoupled, universe was very homogeneous, as showed by the simple relation:
\begin{equation}
\frac{\delta \rho}{\rho} = const \frac{\delta T}{T}
\end{equation}
(\cite{kol}) \footnote{In fact, COBE data gives $\frac{\delta T}{T} \leq 10^{-5}$}. The difference between the actual universe and that at decoupling is evident. The transformation between a highly homogeneous universe, at early times, to an highly local non homogeneous one, can be explained supposing that at $t_{dec}$ were present small inhomogeneities which grow up because of the gravitational instability mechanism (\cite{jea}). Events leading to structure formation can be enumerated as follows:\\
(a) Origin of quantum fluctuations at Planck epoch.\\
(b) Fluctuations enter the horizon and they grow linearly till recombination.\\
(c) Perturbations grow up in a different way for HDM and CDM in the post-recombination phase, till they reach the non-linear phase.\\
(d) Collapse and structure formation.\\ 
Before $ t_{dec} $ inhomogeneities in baryonic components could not grow because photons and baryons were strictly coupled. 
This problem was not present for the CDM component. Then CDM perturbations started to grow up before those in the baryonic component when universe was matter dominated.
The epoch $ t_{eq} \approx 4.4*10^{10}(\Omega_{0} h^{2})^{-2} sec $, at which matter and radiation density are almost equal, can be considered as the epoch at which structures started to form. The study of structure formation is fundamentally an initial value problem. Data necessary for starting this study are:\\
1) Value of $ \Omega_{0} $. In CDM models the value chosen for this parameter is 1, in conformity with inflationary theory predictions.\\
2) The values of $\Omega_i$ for the different components in the universe. For example in the case of baryons, nucleosynthesis gives us the limit $ 0.014 \leq \Omega_{b} \leq 0.15 $ while $ \Omega_{WIMPS} \approx 0.9 $. \\
3) The perturbation spectrum and the nature of perturbations (adiabatic or isocurvature). The spectrum generally used is that of Harrison-Zeldovich: $ P(k) = Ak^{n} $ with $ n=1 $. The perturbation more used are adiabatic or curvature. This choice is  dictated from the comparison between theory and observations of CMBR anisotropy.

\begin{flushleft}
{\it 1.4 The spectrum of density perturbation}
\end{flushleft}


In order to study the distribution of matter density in the universe it is generally assumed that this distribution is given by the superposition of plane waves independently evolving, at least until they are in the linear regime (this means till the overdensity $ \delta = \frac{\rho -\overline{\rho} }{\overline{\rho}}<<1 $). 
Let we divide universe in cells of volume $V_u$ and let we impose periodic conditions on the surfaces. If we indicate with 
$ \overline{\rho} $ the average density in the volume and with $ \rho({\bf r})$ the density in $ {\bf r} $, it is possible to define the density contrast as: 
\begin{equation}
\delta( {\bf r} ) = \frac{\rho( {\bf r}) - \overline{\rho}}{\overline{\rho}}
\end{equation}
This quantity can be developed in Fourier series:
\begin{equation}
\delta({\bf r}) = \sum_{{\bf k}} \delta_{{\bf k}}
exp(i{\bf k} {\bf r}) = \sum_{{\bf k}} \delta_{{\bf k}}
exp(-i{\bf k} {\bf r} ) \label{eq:sovra}
\end{equation}
(\cite{kol}), where $ k_{x} = \frac{2 \pi n_{x}}{l} $ (and similar conditions for the other components) and for the periodicity condition $ \delta(x,y,L) = \delta(x,y,0)$ (and similar conditions for the other components). 
Fourier coefficients $ \delta_{{\bf k}}$ are complex quantities given by: 
\begin{equation}
\delta_{{\bf k}} = \frac{1}{V_{u}} \int_{V_{u}} \delta({\bf r} )
exp(-i{\bf k}{\bf r}) d {\bf r}
\end{equation}
For mass conservation in $V_u$ we have also $ \delta_{{\bf k}=0}=0 $ while for reality of  $ \delta({\bf r}) $,  $ \delta_{{\bf k}}^{\ast} = \delta_{-{\bf k}} $.
If we consider n volumes, $V_u$, we have the problem of determining the distribution of Fourier coefficients $ \delta_{{\bf k}} $ and that of  $ \left|\delta\right| $.
We know that the coefficients are complex quantities and then $ \delta_{{\bf k}} = \left|\delta_{{\bf k}}
\right| exp (i \theta_{{\bf k}})$. If we suppose that phases are random, in the limit $ V_{u} \rightarrow \infty $ 
it is possible to show that  
we get 
$ \left|\delta\right|^{2} =
\sum_{{\bf k}} \left|\delta_{{\bf k}}\right|^{2} $. The Central limit theorem leads us to conclude that the distribution for 
$ \delta $ is Gaussian: 
\begin{equation}
P ( \delta ) \propto exp(\frac{-\delta^{2}}{\sigma^{2}} ) \label{eq:gau}
\end{equation}
(\cite{efs}).
The quantity $\sigma$ that is present in Eq. (\ref{eq:gau}) is the variance of the density field and is defined as: 
\begin{equation}
\sigma^{2} = < \delta ^{2} > = \sum_{{\bf k}} <
\left|\delta_{{\bf k}}\right|^{2} > = \frac{1}{V_{u}}
\sum_{{\bf k}}  \delta_{k}^{2}
\end{equation}
This quantity characterizes the amplitude of the inhomogeneity of the density field. If $ V_{u} \rightarrow \infty$, we obtain the more usual relation:
\begin{equation}
\sigma^{2} = \frac{1}{\left(2 \pi\right)^3} \int P(k) d^{3} k =
\frac{1}{2 \pi^{2}} \int P(k) k^{2} d k
\end{equation}
The term $ P( k ) = < \left|\delta\right|^{2}> $ is called ``Spectrum of perturbations". It is function only of k because the ensemble average in an isotropic universe depends only on r. A choice often made for the primordial spectrum is $ P( k) = A k^{n} $ which in the case $ n = 1 $ gives the scale invariant spectrum of Harrison-Zeldovich. An important quantity connected with the spectrum is the two-points correlation function $ \xi({\bf r}, t) $. It can be defined as the joint probability of finding an overdensity $\delta$ in two distinct points of space:
\begin{equation}
\xi({\bf r}, t) = < \delta( {\bf r},t )
\delta({\bf r}+{\bf x}, t) > \label{eq:corr}
\end{equation}
(\cite{pee1}), where averages are averages on an ensemble obtained from several realizations of universe. Correlation function can be expressed as the joint probability of finding a galaxy in a volume $ \delta V_{1} $ and another in a volume 
$ \delta V_{2} $ separated by a distance $ r_{12} $:
\begin{equation}
\delta ^{2} P = n_{V}^{2} [1+ \xi(r_{12} )] \delta V_{1} \delta V_{2}
\end{equation}
where $ n_{V} $ is the average number of galaxies per unit volume. The concept of correlation function, given in this terms, can be enlarged to the case of three or more points.\\
Correlation functions have a fundamental role in the study of clustering of matter. If we want to use this function for a complete description of clustering, one needs to know the correlation functions of order larger than two (\cite{fry}). By means of correlation functions it is possible to study the evolution of clustering. The correlation functions are, in fact, connected one another by means of an infinite system of equations obtained from moments of Boltzmann equation which constitutes the BBGKY (Bogolyubov-Green-Kirkwood-Yvon) hierarchy (\cite{dav}). 
This hierarchy can be transformed into a closed system of equation using closure conditions. Solving the system one gets information on correlation functions.\\
In order to show the relation between perturbation spectrum and two-points correlation function, we introduce in 
Eq. (\ref{eq:corr}), Eq. (\ref{eq:sovra}), recalling that $ \delta_{{\bf k}}^{\ast} = \delta( -{\bf k} )$ and taking the limit $ V_{u} \rightarrow \infty$, the average in the Eq. (\ref{eq:corr}) can be expressed in terms of the integral: 
\begin{equation}
\xi ( {\bf r} ) = \frac{1}{( 2 \pi ) ^{3}}
\int |\delta( {\bf k} )|^{2} exp( -i {\bf k} {\bf r} ) d^{3} k
\end{equation}
This result shows that the two-point correlation function is the Fourier transform of the spectrum. In an isotropic universe, it is $ |{\bf r}| = r $ and then $ | {\bf k} | =k $ and the spectrum can be obtained from an integral on $ |{\bf k}| = k $. Then correlation function may be written as:
\begin{equation}
\xi( r ) = \frac{1}{2 \pi^{2} }
\int k^{ 2} P (k ) \frac{sin( k r )}{k r} d k
\end{equation}
During the evolution of the universe and after perturbations enter the horizon, the spectrum is subject to modulations because of physical processes characteristic of the model itself (Silk damping (\cite{sil}) for acollisional components, free streaming for collisional particles, etc.). These effects are taken into account by means of the transfer function
$ T(k;t) $ which connects the primordial spectrum $ P( k; t_{p} )$ at time $t_p$ to the final time $t_f$:
\begin{equation}
P(k;t_{f}) = \left[\frac{b(t_{f})}{b(t_{p})}\right]^{2}
T^{2}(k;t_{f}) P( k;t_{p})
\end{equation}
where b(t) is the law of grow of perturbations, in the linear regime. In the case of CDM models the transfer function is:
\begin{equation}
 T( k ) = \left\{1+\left[ak+(bk)^{1.5}+(ck)^{2}
\right]^{\nu}\right\}^{\frac{-1}{\nu}}
\end{equation}
(\cite{bond}), where $a=6.4 (\Omega h^{2} )^{-1}Mpc $; $b=3.0 (\Omega h^{2} )^{-1} Mpc$;
$ c=1.7 (\Omega h^{2} )^{-1} Mpc $; $ \nu=1.13$.
It is interesting to note that Eq. (\ref{eq:gau}) is valid only if $ \sigma << 1 $, since $ \left|\delta\right| \leq 1 $. This implies than non-linear perturbations, $ \sigma >> 1 $, must be non-Gaussian.
In fact when the amplitudes of fluctuations grow up, at a certain point modes are no longer independent and start to couple giving rise to non-linear effects that change the spectrum and correlation function (\cite{jus}).
There are also some theories (e.g., cosmic strings (\cite{kib})) in which even in the linear regime perturbations are not Gaussian. 

\begin{flushleft}
{\it 1.5 Curvature and isocurvature perturbations}
\end{flushleft}


The study of the evolution of density perturbations can be divided into two phases:\\
1) perturbations are outside the horizon, in other terms they have a scale $ \lambda $ larger than Hubble radius $ r_{H} = ct $ or $ \lambda \geq H^{-1} $;\\
2) perturbations are inside the horizon, $\lambda \leq H^{-1} $.\\
In studying the first case it is necessary to use general relativity and one can demonstrate that two different kinds of fluctuations exist: curvature or adiabatic (motivated by the simplest 
models of inflation) and isocurvature or isothermal. Curvature fluctuations are characterized by a fluctuation in energy density or in space curvature. For them, we can write:
\begin{equation}
 \frac{\delta S}{S} = \frac{3}{4} \frac{ \delta \rho_{r}}{\rho_{r}}
-\frac{\delta \rho_{m}}{\delta_{m}} = 3 \frac{\delta T}{T}-
\frac{\delta \rho_{m}}{\rho_{m}} = 0
\end{equation}
(\cite{efs}), where with m has been indicated the matter component, with r radiation, with S entropy and with T temperature. Last equation explains why these fluctuations are named adiabatic, since for them the entropy variation is zero.\\
Isocurvature or isothermal perturbations are not characterized by fluctuations in the curvature of the metric, but they are fluctuations in the local equation of state of universe and in agreement with the name it results $ \delta T = 0$. 
Until fluctuations of isocurvature are not inside the horizon, the causality principle does not permit a redistribution of energy density. This is possible only when perturbations enter the Horizon and isocurvature perturbations can be converted into perturbations in the energy density. As a consequence, the distinction between those two kinds of perturbations is no longer meaningful after them enter the horizon (\cite{sut}; \cite{gou}).
The origin of curvature perturbations may be explained inside inflationary models or assuming that they are initially present as perturbations of the metric. Isocurvature fluctuations may be always produced in inflationary scenarios from fluctuations in the density number of barions or axions (\cite{efs}, \cite{kol}).

\begin{flushleft}
{\it 1.6 Perturbations evolution}
\end{flushleft}


Density perturbations in the components of the universe evolve with time. In order to get the evolution equations for $\delta$ in Newtonian regime, it is possible to use several models. In our model, we assume that gravitation dominates on the other interactions and that particles (representing galaxies, etc.) move collisionless in the potential $\phi$ of a smooth density function (\cite{pee1}). The distribution function of particles for position and momentum is given by:
\begin{equation}
d N = f( {\bf x}, {\bf p},t) d^{3}x d^{3} p
\end{equation}
and density:
\begin{equation}
\rho({\bf x} , t )= m a^{-3} \int d^{3} p f({\bf x}, {\bf p}, t)=
\rho_{b}\left[1+ \delta({\bf x}, t)\right] \label{eq:densi}
\end{equation}
where m is the mass of a particle and $\rho_{b}$ the background density. Applying Liouville theorem to the probability density on a limited region of phase-space of the system we have that f verifies the equation:
\begin{equation}
\frac{\partial f}{\partial t }+\frac{{\bf p}}{m a^{2}}
\bigtriangledown f - m \bigtriangledown \phi \frac{
\partial f}{\partial {\bf p}} = 0 \label{eq:liou}
\end{equation}
The distribution function f that appears in the previous equations cannot be obtained from observations. It is possible to measure moments of f (density, average velocity, etc.). We want now to obtain the evolution equations for $\delta$. For this reason, we start integrating Eq. (\ref{eq:liou}) on $ {\bf p} $ and after using Eq. (\ref{eq:densi}), we get:
\begin{equation}
a^{3} \rho_{b} \frac{\partial \delta}{\partial t}+
\frac{1}{a^{2}} \bigtriangledown \int {\bf p} f d^{3} p =0 \label{eq:tre}
\end{equation}
If we define velocity as: 
\begin{equation}
{\bf v} = \frac{\int \frac{{\bf p}}{m a}f d^{3} p}{\int f d^{3} p}
\end{equation}
and introduce it in Eq. (\ref{eq:tre}) we get:
\begin{equation}
\rho_{b} \frac{\partial \delta}{\partial t}+
\frac{1}{a} \bigtriangledown (\rho {\bf v} )= 0
\end{equation}
We can now multiply Eq. (\ref{eq:liou}) for $ {\bf p} $ and integrate it on the momentum:
\begin{equation}
\frac{\partial}{\partial t} \int p_{\alpha} f d^{3} p +
\frac{1}{m a^{2}} \partial_{\beta} \int p_{\alpha}
p_{\beta} f d^{3} p + a^{3} \rho ( {\bf x}, t) \phi_{, \alpha} =0
\end{equation}
this last in Eq. (\ref{eq:tre}) leaves us with:
\begin{equation}
\frac{\partial^{2} \delta }{\partial t^{2}} + 2\frac{\dot{a}}{a}
\frac{\partial \delta}{\partial t} = \frac{1}{a^{2}}
\bigtriangledown \left[(1+\delta ) \bigtriangledown \phi\right]+
\frac{1}{\rho_{b} m a^{7}} \partial_{\alpha} \partial_{\beta}
\int p_{\alpha} p_{\beta} \phi d^{3} p
\end{equation}
and finally using 
\begin{equation}
< v_{\alpha} v_{\beta} > = \frac{\int f p_{\alpha} p_{\beta} d^{3} p }
{m a^{2} \int f d^{3} p}
\end{equation}
the equation for the evolution of overdensity becomes:
\begin{equation}
\frac{\partial^{2} \delta }{\partial t^{2}} + 2\frac{\dot{a}}{a}
\frac{\partial \delta}{\partial t} = \frac{1}{a^{2}}
\bigtriangledown \left[(1+\delta ) \bigtriangledown \phi\right]+
\frac{1}{a^{2}} \partial_{\alpha} \partial_{\beta}
\left[(1+\delta) < v^{\alpha} v^{\beta} >\right]
\end{equation}
(\cite{pee1}). The term $ < v_{\alpha} v_{\beta} > $ is the tensor of anisotropy of peculiar velocity. This is present in the gradient and then it behaves as a pressure force. If we consider an isolated and spherical perturbation, it is possible to assume that initial asymmetries does not grow up and so we can suppose, in this hypothesis that $ < v_{\alpha} v_{\beta} > = 0 $. In this case and with the linearity assumption $ \delta << 1 $ we have:
\begin{equation}
\frac{\partial^{2} \delta}{\partial t^{2}} + 2\frac{\dot{a}}{a}
\frac{\partial \delta}{\partial t } = 4 \pi G \rho_{b} \delta
\end{equation}
This equation in an Einstein-de Sitter universe ($\Omega = 1$,
$\Lambda = 0 $) has the solutions:
\begin{equation}
\delta_{+}= A_{+}({\bf x}) t^{\frac{2}{3}} \hspace{1cm}
\delta_{-} ({\bf x},t) = A_{-} ({\bf x}) t^{-1}
\end{equation}
The perturbation is then done of two parts: a growing one, becoming more and more important with time, and a decaying one
becoming negligible with increasing time, with respect to the growing one. 

In the case of open models with no cosmological constant: $\Omega<1$,
$\Lambda=0$, we can write:

\begin{equation} 
\frac{\dot{a}^2}{a^2}=\frac{8}{3}\pi G\bar{\rho}\left(1+
\left(\Omega_0^{-1} -1\right) a\right), \label{eq:frw_open} 
\end{equation}
\noindent 
and the $a(t)$ evolution can be expressed through the following
parametric representation:

\begin{eqnarray} 
a(\eta)&=&\frac{\Omega_0}{2(1-\Omega_0)}({\rm cosh} \eta -1) \\
     t(\eta)&=&\frac{\Omega_0}{2H_0(1-\Omega_0)^{3/2}}({\rm sinh}\eta -\eta).
\nonumber \label{eq:a_ev_open} 
\end{eqnarray} 
In the case of flat models with positive cosmological constant: $\Omega<1$,
$\Lambda\neq 0$, $\Omega+\Lambda/3H_0^2 = 1$, we can write:
\begin{equation} 
\frac{\dot{a}^2}{a^2}=\frac{8}{3}\pi G\bar{\rho}\left(1+
\left(\Omega_0^{-1} -1\right) a^3\right), \label{eq:frw_cosm} 
\end{equation} 
\begin{equation} 
a(t) = \left(\Omega_0^{-1}-1\right)^{-1/3}{\rm sinh}^{2/3}
\left(\frac{3}{2}\sqrt{\frac{\Lambda}{3}} t\right). \label{eq:a_ev_cosmo} 
\end{equation} 

Before concluding this section, we want to find an expression for the velocity field in the linear regime. 
Using the equation of motion $ {\bf p} = m a^{2} {\bf \dot{x}} $, $ \frac{d {\bf p}}{d t} = -m \bigtriangledown \phi $ and the proper velocity of a particle, $ v = a \bf{\dot{x}} $, verify the equation:
\begin{equation}
\frac{d {\bf v}}{d t} + {\bf v} \frac{\dot{a}}{a} =
-\frac{\bigtriangledown \phi}{a} = G \rho_{b} a
\int d^{3} x \delta({\bf x}, t) \frac{{\bf x} -{\bf x'}}{\left|{\bf x}-
{\bf x'}\right|}
\end{equation}
Supposing that ${\bf v} $ is a similar solution for the density, $ {\bf v} = {\bf V_{+}({\bf x}, t)} t^{p}$, we get:
\begin{equation}
v^{\alpha} = \frac{H a}{4 \pi} \frac{\partial}{\partial x^{\alpha}}
\int d^{3} x' \frac{\delta ( {\bf x'}, t)}{\left|{\bf x'}-
{\bf x}\right|}
\end{equation}
(\cite{pee1}). 
This solution is valid just as that for $ \delta $ in the linear regime. At time $ t = t_{0} $ this regime is valid on scales larger than $ 8 h^{-1} Mpc $.

\begin{flushleft}
{\it 1.7 Non-linear phase}
\end{flushleft}


Linear evolution is valid only if $ \delta << 1 $ or similarly, if the mass variance, $\sigma$, is much less than    
unity. When this condition is no longer verified (e.g., if we consider scales smaller than $8h^{-1}$ Mpc), it is necessary to develope a non-linear theory. In regions smaller than $8h^{-1}$ Mpc
galaxies are not a Poisson distribution but they tend to cluster.   
If one wants to study the properties of galactic structures or clusters of galaxies, it is necessary to introduce a non-linear theory of clustering. A theory of this last item is too complicated to be developed in a purely theoretical fashion. The problem can be faced assuming certain approximations that simplifies it (\cite{zel}) or as often it is done, by using N-Body simulations of the interesting system.  
The approximations are often used to furnish the initial data to simulations. 
In the simulations, a large number of particles are randomly distributed in a sphere, in the points of a cubic grid, in order to eliminate small scale noise. The initial spectrum is obtained perturbing the initial positions by means of a superposition of plane waves having random distributed phases and wave vector (\cite{wes}).
Obviously, the universe is considered in expansion (or comoving coordinates are used), and then the equation of motion of particles are numerically solved. For what concerns the analytical approximations one of the most used is that of \cite{zel}. This gives a solution to the problem of the grow of perturbations in an universe with $p=0$ not only in the linear regime but even in the mildly non-linear regime.
In this approximation, one supposes to have particles with initial position given in Lagrangian coordinates $ {\bf q} $.
The positions of particles, at a given time t, are given by: 
\begin{equation}
{\bf x}={\bf q}+b(t){\bf p(q)}
\end{equation}
where ${\bf x}$ indicates the Eulerian coordinates, ${\bf p(q) } $ describes the initial density fluctuations and $b(t)$ describes their grow in the linear phase and it satisfies the equation:
\begin{equation}
\frac{d^{2}b}{dt^{2}}+2a^{-1}\frac{db}{dt}\frac{da}{dt}=4 \pi G
\rho b
\end{equation}
The equation of motion of particles, according to the quoted approximation, is given by: 
\begin{equation}
{\bf v} = \dot{a} {\bf q} + \dot{b} {\bf p}({\bf q} )
\end{equation}
The peculiar velocity of particles is given by:
\begin{equation}
{ \bf u}=\frac{d{\bf x}}{dt} =\frac{db}{dt}{ \bf p(q)}
\end{equation}
while the density of the perturbed system is given by:
\begin{equation}
\rho({\bf q},t)=\overline{\rho}
\left|\frac{\partial q_{j}}{\partial x_{k}}\right| =
\overline {\rho} \left| \delta_{jk} + b(t)
\frac{\partial p_{k} }{\partial q_{j} }\right|^{-1} \label{eq:cas}
\end{equation}
Developing the Jacobian present in Eq. (\ref{eq:cas}) at first order in $ b(t) {\bf p(q)} $, one obtains:
\begin{equation}
\frac{\delta \rho}{\overline{\rho}}\approx -b(t)
\bigtriangledown_{{ \bf q}}{\bf p(q)}
\end{equation}
This equation can be re-written, separating the space and time dependence, as in the equation for $ { \bf u} $, and writing:
\begin{equation}
b(t)=t^{\frac{2}{3}} \hspace{1.0cm} {\bf p(q)}=
\sum_{{\bf k}}i\frac{{\bf k}}{\left|{\bf k}\right|^{2}}A_{{\bf k}}
exp(i{\bf k q})
\end{equation}
in the form: 
\begin{equation}
\frac{\delta \rho}{\overline{\rho}}=
\sum_{{ \bf k}} A_{{\bf k}}t^\frac{2}{3} exp(i{\bf k q})
\end{equation}
(\cite{efs}), that leads us back to the linear theory. 
In other words, Ze`ldovich approximation is able to reproduce the linear theory, and is also able to give a good approximation in regions with $ \frac{\delta \rho}{\overline{\rho}}>>1$. Using the expression for $ p(q) $, the Jacobian in Eq. (\ref{eq:cas}) is a real matrix and symmetric that can be diagonalized. 
With this $ p(q) $ the perturbed density can be written as:
\begin{equation}
\rho({\bf q},t)=\frac{\overline{\rho}}{(1-b(t)\lambda_{1}(q))
(1-b(t)\lambda_{2}(q))(1-b(t)\lambda_{3}(q))} \label{eq:panc}
\end{equation}
where $ \lambda_{1} $, $ \lambda_{2} $, $ \lambda_{3} $ are the three eigenvalues of the Jacobian, describing the expansion and contraction of mass along the principal axes. From the structure of the last equation, we notice that in regions of high density Eq. (\ref{eq:panc}) becomes infinite and the structure of collapse in a pancake, in a filamentary structure or in a node, according to values of eigenvalues. Some N-body simulations (\cite{efs1}) tried to verify the prediction of 
Ze`ldovich approximation, using initial conditions generated using a spectrum with a cut-off at low frequencies. 
The results showed a good agreement between theory and simulations, for the initial phases of the evolution 
($ a(t)=3.6 $). Going on, the approximation is no more valid starting from the time of shell-crossing. After shell-crossing, particles does not oscillate any longer around the structure but they pass through it making it vanish. 
This problem has been partly solved supposing that particles, before reaching the singularity they sticks the one on the other, due to a dissipative term that simulates gravity and then collects on the forming structure.  
This model is known as ``adesion-model" (\cite{gur}). \\
Summarizing, Zel'dovich approximation gives a description of the transition between linear and non-linear phase. It is expecially used to get the initial conditions for N-body simulations. 

\begin{flushleft}
{\it 1.8 Quasi-linear regime}
\end{flushleft}


We have seen in the previous section that in the case of regions of dimension smaller than $ 8 h^{-1} Mpc $, the linear theory is no more a good approximation and  a new theory is needed or  N-body simulations. 
Non-linear theory is able to calculate quantities as the formation redshift of a given class of objects as galaxies and clusters, the number of bound objects having masses larger than a given one, the average virial velocity and the correlation function. It is possible to get an estimate of the given quantities as that of other not cited, using an intermediate theory between the linear and non-linear theory: the quasi-linear theory. This last is obtained adding to the linear theory a model of gravitational collapse, just as the spherical collapse model. Important results that the theory gives is the bottom-up formation of structures (in the CDM model). Other important results are obtained if we identify density peaks in linear regime with sites of structure formation. Two important papers in the development of this theory are \cite{pre} and that of \cite{bar}. This last paper is an application of the ideas of the quasi-linear theory to the CDM model. The principles of this approach are the following:\\
\begin{itemize}
\item
Regions of mass larger than M that collapsed can be identified with regions where the density contrast evolved according to linear regime, $ \delta ( M, x ) $, has a value larger than   a threshold, $ \delta_{c} $.
\item After collapse regions does not fragment.
\end{itemize}
The major drawbacks of the theory, as described in \cite{bar} are fundamentally the fact that the estimates that can be obtained by means of this theory depends on the threshold $ \delta_{c} $, on the ratio between the filtering mass and that of objects and from other parameters. Nevertheless, this theory has helped cosmologists in obtaining estimate of important quantities as those previously quoted, and at same time give evidences that leads to exclude very low values for spectrum normalization.

\begin{flushleft}
{\it 1.9 Spherical Collapse}
\end{flushleft}


Spherical symmetry is one of the few cases in which gravitational
collapse can be solved exactly (\cite{gun}; \cite{pee1}).  In
fact, as a consequence of Birkhoff's theorem, a spherical perturbation
evolves as a FRW Universe with density equal to the mean density
inside the perturbation.

The simplest spherical perturbation is the top-hat one, i.e. a
constant overdensity $\delta$ inside a sphere of radius $R$; to avoid
a feedback reaction on the background model, the overdensity has to be
surrounded by a spherical underdense shell, such to make the total
perturbation vanish. The evolution of the radius of the perturbation
is then given by a Friedmann equation.

The evolution of a spherical perturbation depends only on its initial
overdensity. In an Einstein-de Sitter background, any spherical 
overdensity
reaches a singularity (collapse) at a final time:

\begin{equation} t_c=\frac{3\pi}{2}\left(\frac{5}{3}\delta(t_i)\right)^{-3/2} t_i.
\label{eq:spherical_coll} \end{equation}

\noindent
By that time its linear density contrast reaches the value: 

\begin{equation} \delta_l(t_c)=\delta_c=\frac{3}{5}\left(\frac{3\pi}{2}\right)^{3/2}
\simeq 1.69. \label{eq:delta_c_sph}\end{equation}

\noindent
In an open Universe not any overdensity is going to collapse: the
initial density contrast has to be such that the total density inside
the perturbation overcomes the critical density. This can be
quantified (not exactly but very accurately) as follows: the growing
mode saturates at $b(t)=5/2(\Omega_0^{-1}-1)$, so that a perturbation
ought to satisfy $\delta_l>1.69\cdot 2(\Omega_0^{-1}-1)/5$ to be able
collapse.

Of course, collapse to a singularity is not what really happens in
reality. It is typically supposed that the structure reaches virial
equilibrium at that time. In this case, arguments based on the virial
theorem and on energy conservation show that the structure reaches a
radius equal to half its maximum expansion radius, and a density
contrast of about 178. In the subsequent evolution the radius and the
physical density of the virialized structure remains constant, and its
density contrast grows with time, as the background density decays.
Similarly, structures which collapse before are denser than the ones
which collapse later.

Spherical collapse is not a realistic description of the formation of
real structures; however, it has been shown (see \cite{ber} for
a rigorous proof or \cite{val1}, \cite{val2}) that high peaks ($> 2\sigma$) follow spherical
collapse, at least in the first phases of their evolution. However, 
a small systematic departure from
spherical collapse can change the statistical properties of collapse
times.

Spherical collapse can describe the evolution of underdensities. A
spherical underdensity is not able to collapse (unless the Universe is
closed!), but behaves as an open Universe, always expanding unless its
borders collide with neighboring regions. At variance with
overdensities, underdensities tend to be more spherical as they
evolve, so that the spherical model provides a very good approximation
for their evolution.

\begin{flushleft}
{\it 1.9.1 Improvements to the Spherical Collapse model}
\end{flushleft}


Several years ago it was realized that the density field distributions
around the density peaks, which eventually will give birth to galaxies and
clusters, depart from spherical symmetry and from the average density
profile, producing important consequences on collapse dynamics and formation
of protostructures (\cite{hofsha}; \cite{ryd}; \cite{ryd1}; 
\cite{kashlinsky}, \cite{kashlinsky1}; \cite{peebles90}). A fundamental role in this
context is played by the joint action of tidal torques (coupling shells of
matter which are accreted around a density peak and neighboring
protostructures (\cite{ryd}; \cite{ryd1})), and by dynamical friction (\cite{white76};
\cite{kashlinsky}, \cite{kashlinsky1}, \cite{ac}).

According to the previrialization conjecture 
(\cite{davispe}, \cite{peebles90}), initial asphericities and tidal interactions between neighboring
density fluctuations induce significant non-radial motions which oppose the
collapse. This means that virialized clumps form later, with respect to the
predictions of the linear perturbation theory or the spherical collapse model (hereafter SM),
and that the initial density contrast, needed to obtain a given final
density contrast, must be larger than that for an isolated spherical
fluctuation.
This kind of conclusion was supported by \cite{barrowsi}, \cite{szasi}, \cite{villda}, \cite{bond2} 
and \cite{lokju}. \\
In particular \cite{barrowsi} and \cite{szasi} pointed out
that non-radial motions would slow the rate of growth of the density contrast
by lowering the peculiar velocity and suppress collapse once the system
detaches from general expansion. \cite{villda} gave examples
of the growth of non-radial motions in N-body simulations.
Arguments based on a numerical least-action method lead \cite{peebles90}
to the conclusion that irregularities in the mass distribution,
together with external tides, induce
non-radial motions
that slow down the collapse.
\cite{lokju} used N-body simulations and a weakly non-linear
perturbative approach to study previrialization. They concluded that when
the slope of the initial power spectrum is $n>-1$, non-linear tidal
interactions slow down the growth of density fluctuations and the
magnitude of the effect increases when $n$ is increased. \\
Opposite conclusions were obtained by \cite{hofma},
\cite{evrard}, \cite{beja}, \cite{monaco}.
In particular \cite{hofma}, using the quasi-linear (QL) approximation
(\cite{zel}) showed that the shear affects
the dynamics of collapsing objects and it leads to infall velocities that are
larger than in the case of non-shearing ones. Bertschinger \& Jain (\cite{beja}) put
this result in theorem form, according to which spherical perturbations are
the slowest in collapsing. 
Bartelmann et al. (1993) argued that the collapse does not start from a comoving motion of the perturbation, but that the continuity 
equation requires an initial velocity perturbation directly related to the density perturbation. The effect is that collapse proceeds faster than in the case where the initial velocity perturbation is set equal to zero and the collapse timescale is shortened. 
The N-body simulations by \cite{evrard}
did not reproduce previrialization effect, but the reason is due to the fact
that they assumed an $n=-1$ spectrum, differently from the $n=0$ one used by
\cite{peebles90} that reproduced the effect. If $n<-1$ the peculiar gravitational
acceleration, $g \propto R^{-(n+1)/2}$, diverges at large $R$ and the
gravitational acceleration moves the fluid more or less uniformly,
generating bulk flows rather than shearing motions. Therefore, its collapse
will be similar to that of an isolated spherical clump. If $n>-1$, the
dominant sources of acceleration are local, small-scale inhomogeneities and
tidal effects will tend to generate non-radial motions and resist gravitational
collapse.
In a more recent paper, \cite{audit} have
proposed some analytic prescriptions to compute
the collapse time along the second and the third principal axes of an
ellipsoid,
by means of the 'fuzzy'  threshold approach.
They pointed out that the formation of real virialized clumps must correspond
to the third axis collapse and that the collapse along this axis
is slowed down by the  effect of the shear
rather than be accelerated by it,
in contrast to its effect on the first axis collapse.
They concluded that spherical collapse is the fastest, in disagreement with
Bertschinger \& Jain's theorem.
This result
is in agreement with \cite{peebles90}.
The quoted controversy was addressed by 
\cite{del3} who examined the evolution of non-spherical inhomogeneities in a Einstein-de
Sitter universe, by numerically solving the equations of motion for the principal
axes and the density of a dust ellipsoid.
%
%
They showed that for lower values of $\nu$ ($\nu=2$) the
growth rate enhancement of the density contrast induced by 
the shear is counterbalanced by the effect of angular momentum acquisition.
For $\nu>3$ the effect of angular momentum and shear reduces, and the
evolution of perturbations tends to follow the behavior obtained in the SM.
\cite{del2002} studied the role of shear fields on the evolution of density perturbations
by using an analytical approximate solution for the equations of
motion of homogeneous ellipsoids embedded in a homogeneous
background. The equations of motion of a homogeneous ellipsoid (\cite{icke};
\cite{whsi}(hereafter WS)) were modified in order to take
account of the tidal field, as done in \cite{watanabe} and then were
integrated analytically, similar to what was done in WS. 
The density contrast at turn-around and the
collapse velocity were found to be reduced with respect to that
found by means of the SM. The reduction increases
with increasing strength of the external tidal field and with
increasing initial asymmetry of the ellipsoids. 

The second physical effect with changes cluster collapse is dynamical friction. Former treatments of the dynamical friction effects on the structure of clusters of galaxies, considering only the component generated 
by the galactic population on the motion of galaxies themselves are due to \cite{white76} and \cite{kash}; \cite{kashlinsky}; \cite{kashlinsky1}. AC
recalculated the effect of dynamical friction taking into account the effect of substructure, showing that
dynamical friction delays the collapse of low-$\nu $ peaks
inducing a bias of dynamical nature. 
Because of dynamical friction
under-dense regions in clusters (the clusters outskirts) accrete less mass
with respect to that accreted in absence of this dissipative effect and as a
consequence over-dense regions are biased toward higher mass (AC).
Dynamical friction and non-radial motions acts
in a similar fashion: they delay the shell collapse
consequently inducing a dynamical bias.
Whenever efficient, these mechanisms will generate a physical
selection of those peaks in the initial density field that eventually will
give rise to the observed cosmic structures. 
As a consequence of dynamical friction and tidal torques, one expects changes in the threshold of collapse, the mass function and the correlation function.

In the next subsections, I shall study how the spherical collapse model is changed by the joint effect of 
dynamical friction, tidal torques and a non-zero cosmological constant.

\begin{flushleft}
{\it 1.9.2 Dynamical friction and structure formation}
\end{flushleft}


In a hierarchical structure formation model, the large scale cosmic
environment can be represented as a collisionless medium made of a hierarchy
of density fluctuations whose mass, $M$, is given by the mass function $%
N(M,z)$, where $z$ is the redshift. In these models matter is concentrated
in lumps, and the lumps into groups and so on.
In such a material system, gravitational
field can be decomposed into an average field, ${\bf F}_0(r)$, generated
from the smoothed out distribution of mass, and a stochastic component, $%
{\bf F}_{stoch}(r)$, generated from the fluctuations in number of the field
particles. 
The stochastic component of the gravitational field is specified assigning a
probability density, $W({\bf F})$, (\cite{chandra}). In
an infinite homogeneous unclustered system $W({\bf F})$ is given by
Holtsmark distribution (\cite{chandra}) while in
inhomogeneous and clustered systems $W({\bf F})$ is given by \cite{kandrup}
and \cite{anba} respectively. The stochastic
force, ${\bf F}_{stoch}$, in a self-gravitating system modifies the motion
of particles as it is done by a frictional force. In fact a particle moving
faster than its neighbors produces a deflection of their orbits in such a
way that average density is greater in the direction opposite to that of
traveling causing a slowing down in its motion. Following \cite{chandra} method, the frictional force which is experienced by
a body of mass $M$ (galaxy), moving through a homogeneous and isotropic
distribution of lighter particles of mass $m$ (substructure), having a
velocity distribution $n(v)$ is given by: 
\begin{equation}
M\frac{d{\bf v}}{dt}=-4\pi G^2M^2n(v)\frac{{\bf v}}{v^3}\log \Lambda \rho 
\label{eq:cha}
\end{equation}
where $\log \Lambda $ is the Coulomb logarithm, $\rho $ the density of the
field particles (substructure). \\ A more general formula is that given by
\cite{kandrup} in the hypothesis that there are no correlations among random
force and their derivatives: 
\begin{equation}
{\bf F}=-\eta {\bf v}=-\frac{\int W(F)F^2T(F)d^3F}{2<v^2>}{\bf v}
\end{equation}
where $\eta $ is the coefficient of dynamical friction, $T(F)$ the average
duration of a random force impulse, $<v^2>$ the characteristic speed of a
field particle having a distance $r\simeq (\frac{GM}F)^{1/2}$ from a test
particle (galaxy). This formula is more general than Eq. (\ref{eq:cha})
because the frictional force can be calculated also for inhomogeneous
systems when $W(F)$ is given. If the field particles are distributed
homogeneously the dynamical friction force is given by: 
\begin{equation}
F=-\eta v=-\frac{4.44G^2m_a^2n_a}{[<v^2>]^{3/2}}\log \left\{ 1.12\frac{<v^2>%
}{Gm_an_a^{1/3}}\right\} 
\end{equation}
(\cite{kandrup}), where $m_a$ and $n_a$ are respectively the average mass and
density of the field particles. Using virial theorem we also have: 
\begin{equation}
\frac{<v^2>}{Gm_an_a^{1/3}}\simeq \frac{M_{tot}}m\frac
1{n^{1/3}R_{sys}}\simeq N^{2/3}
\end{equation}
where $M_{tot}$ is the total mass of the system, $R_{sys}$ its radius and $N$
is the total number of field particles. The dynamical friction force can be
written as follows: 
\begin{equation}
F=-\eta v=-\frac{4.44[Gm_an_{ac}]^{1/2}}N\log \left\{ 1.12N^{2/3}\right\}
\frac v{a^{3/2}}=-\epsilon_o \frac v{a^{3/2}}
\end{equation}
where $N=\frac{4\pi }3R_{sys}^3n_a$ and $n_{ac}=n_a\times a^3$ is the
comoving number density of peaks of substructure of field particles. This
last equation supposes that the field particles generating the stochastic
field are virialized. This is justified by the previrialization hypothesis
(\cite{davispe}). \\ To calculate the dynamical evolution of the
galactic component of the cluster it is necessary to calculate the number
and average mass of the field particles generating the stochastic field. \\ %
The protocluster, before the ultimate collapse at $z\simeq 0.02$, is made of
substructure having masses ranging from $10^6-10^9M_{\odot }$ and from
galaxies. I suppose that the stochastic gravitational field is generated
from that portion of substructure having a central height $\nu $ larger than
a critical threshold $\nu _c$. This latter quantity can be calculated
(following AC) using the condition that the peak radius, $r_{pk}(\nu \ge \nu
_c),$ is much less than the average peak separation $n_a(\nu \ge \nu
_c)^{-1/3}$, where $n_a$ is given by the formula of \cite{bar} for the upcrossing
points: 
\begin{eqnarray}
n_{ac}(\nu \ge \nu _c)=\frac{\exp (\nu _c^2/2)}{(2\pi )^2}(\frac \gamma
{R_{*}})^3 [ \nu _c^2-1+ \nonumber \\
\frac{4\sqrt{3}}{5\gamma ^2(1-5\gamma ^2/9)^{1/2}}
\exp ( -5\gamma ^2\nu _c^2/18)] 
\end{eqnarray}
where $\gamma $, $R_{*}$ are parameters related to moments of the power
spectrum (\cite{bar} Eq. ~4.6A). The condition $r_{pk}(\nu \ge \nu _c)<0.1n_a(\nu
\ge \nu _c)^{-1/3}$ ensures that the peaks of substructure are point like.
Using the radius for a peak: 
\begin{equation}
r_{pk}=\sqrt{2}R_{*}\left[ \frac 1{(1+\nu \sigma _0)(\gamma ^3+(0.9/\nu
))^{3/2}}\right] ^{1/3}
\end{equation}
(AC), I obtain a value of $\nu _c=1.3$ and then we have $n_a(\nu \ge \nu
_c)=50.7Mpc^{-3}$ 
($\gamma =0.4$, $R_{*}=50Kpc$) and $m_a$ is given by: 
\begin{equation}
m_a=\frac 1{n_a(\nu \ge \nu _c)}\int_{\nu _c}^\infty m_{pk}(\nu )N_{pk}(\nu
)d\nu =10^9M_{\odot }
\end{equation}
(in accordance with the result of AC), where $m_{pk}$ is given in \cite{pehe} and $N_{pk}$ is the average number density of peak (\cite{bar}
Eq. ~4.4).
Galaxies and Clusters of galaxies are correlated systems whose autocorrelation function, $\xi(r)$,
can be expressed, 
in a power law form (\cite{pee1}; \cite{bah1}; \cite{postman}; \cite{davispe1}; \cite{gonzalez}).
%
%
The description of dynamical friction in these systems need to
use a distribution of the stochastic forces, $W(F)$, taking account of
correlations. In this last case the coefficient of dynamical friction,
$ \eta$, may be calculated using the equation:
\begin{equation}
\eta=\int  d^{3} {\bf F} W(F) F^{2} T(F)/(2<v^{2}>)
\end{equation}
and using \cite{anba} distribution:
\begin{equation}
W(F)=\frac{1}{2 \pi^2 F} \int_{0}^{\infty} dk k sin(kF)A_{f}(k)
\end{equation}
where $A_{f}$, which is a linear integral function of the correlation function $ \xi(r)$,
is given in the quoted paper (Eq. 36). 

\begin{flushleft}
{\it 1.9.3 Tidal torques and structure formation}
\end{flushleft}


The explanation of galaxies spins gain through tidal torques was pioneered
by \cite{hoyle}. Peebles (\cite{peeb69}) performed the first detailed calculation of the
acquisition of angular momentum in the early stages of protogalactic
evolution. More recent analytic computations (\cite{whi84}, \cite{hofma},
\cite{ryd}; \cite{eilo}; \cite{catel}; \cite{catel1} and numerical simulations (\cite{barnef}) have
re-investigated the role of tidal torques in originating galaxies angular
momentum. 
One way to study the variation of angular momentum with radius in
a galaxy is that followed by \cite{ryd}. In this approach the protogalaxy
is divided into a series of mass shells and the torque on each mass shell is
computed separately. The density profile of each proto-structure is
approximated by the superposition of a spherical profile, $\delta (r)$, and
a random CDM distribution, ${\bf \varepsilon (r)}$, which provides the
quadrupole moment of the protogalaxy. 
As shown by \cite{ryd} the net rms torque on a
mass shell centered on the origin of internal radius $r$ and thickness $%
\delta r$ is given by: 
\begin{eqnarray}
\langle |\tau |^2\rangle ^{1/2}=\sqrt{30}\left( \frac{4\pi }5G\right) 
[\langle a_{2m}(r)^2\rangle \langle q_{2m}(r)^2\rangle \nonumber \\
-\langle a_{2m}(r)q_{2m}^{*}(r)\rangle ^2] ^{1/2}  \label{eq:tau}
\end{eqnarray}
where $q_{lm}$, the multipole moments of the shell and $a_{lm}$, the tidal
moments, are given by: 
\begin{equation}
\langle q_{2m}(r)^2\rangle =\frac{r^4}{\left( 2\pi \right) ^3}M_{sh}^2\int
k^2dkP\left( k\right) j_2\left( kr\right) ^2
\end{equation}
\begin{equation}
\langle a_{2m}(r)^2\rangle =\frac{2\rho _b^2r^{-2}}\pi \int dkP\left(
k\right) j_1\left( kr\right) ^2
\end{equation}
\begin{equation}
\langle a_{2m}(r)q_{2m}^{*}(r)\rangle =\frac r{2\pi ^2}\rho _bM_{sh}\int
kdkP\left( k\right) j_1\left( kr\right) j_2(kr)
\end{equation}
where $M_{sh}$ is the mass of the shell, $j_1(r)$ and $j_2(r)$ are the
spherical Bessel function of first and second order while the power spectrum 
$P(k)$ is given by, \cite{bar} (equation~(G3)): 
\begin{eqnarray}
T(k) &=& \frac{[\ln \left( 1+2.34 q\right)]}{2.34 q}
\nonumber \\
& & 
\cdot [1+3.89q+
(16.1 q)^2+(5.46 q)^3+(6.71)^4]^{\frac{-1}{4}}
%
%
\label{eq:ma5}
\end{eqnarray}
(where 
$q=\frac{k\theta^{1/2}}{\Omega_{\rm X} h^2 {\rm Mpc^{-1}}}$.
Here $\theta=\rho_{\rm er}/(1.686 \rho_{\rm \gamma})$
represents the ratio of the energy density in relativistic particles to
that in photons ($\theta=1$ corresponds to photons and three flavors of
relativistic neutrinos).
The power spectrum was normalized to reproduce the observed abundance of rich 
cluster of galaxies.
Filtering the spectrum on cluster scales, $R_f=3h^{-1}Mpc$, I have obtained
the rms torque, $\tau (r)$, on a mass shell using Eq. (\ref{eq:tau}) then I
obtained the total specific angular momentum, $h(r,\nu )$, acquired during
expansion integrating the torque over time (\cite{ryd} Eq. 35): 
\begin{eqnarray}
h(r,\nu )=\frac 13\left( \frac 34\right) ^{2/3} \nonumber \\
\frac{\tau _ot_0}{M_{sh}}%
\overline{\delta }_o^{-5/2}\int_0^\pi \frac{\left( 1-\cos \theta \right) ^3}{%
\left( \vartheta -\sin \vartheta \right) ^{4/3}}\frac{f_2(\vartheta )}{%
f_1(\vartheta )-f_2(\vartheta )\frac{\delta _o}{\overline{\delta _o}}}%
d\vartheta   \label{eq:ang}
\end{eqnarray}
the functions $f_1(\vartheta )$, $f_2(\vartheta )$ are given by \cite{ryd} (Eq. 31) while the mean over-density inside the shell, $\overline{%
\delta }(r)$, is given by \cite{ryd}: 
\begin{equation}
\overline{\delta }(r,\nu )=\frac 3{r^3}\int_0^\infty d\sigma \sigma ^2\delta
(\sigma )
\label{eq:denss}
\end{equation}
where $\delta(r)=\frac{\rho(r)-\rho_b}{\rho_b}$.
%
%
As showed by \cite{ryd}, the rms specific angular momentum, 
$h(r,\nu )$, increases with distance $r$ while peaks of greater $\nu $
acquire less angular momentum via tidal torques. This is the angular
momentum-density anticorrelation showed by \cite{hofma}. This effect
arises because the angular momentum is proportional to the gain at turn
around time, $t_m$, which in turn is proportional to $\overline{\delta }%
(r,\nu )^{-\frac 32}\propto \nu ^{-3/2}$.

\begin{flushleft}
{\it 1.9.4 Modification of collapse}
\end{flushleft}


Tidal torques and dynamical friction acts in a similar fashion. As previously reported, 
AC calculated the effect of dynamical friction taking into account the effect of substructure, showing that
dynamical friction delays the collapse of low-$\nu $ peaks
inducing a bias of dynamical nature. Similarly non-radial motions would slow the rate of growth of the density contrast
by lowering the peculiar velocity and suppress collapse once the system
detaches from general expansion. 
In fact, in the central regions of a density peak ($r\leq 0.5R_f$) the velocity
dispersion attain nearly the same value 
while at larger radii ($r\geq R_f$) the radial component is lower than the
tangential component. This means that motions in the outer regions are
predominantly non-radial and in these regions the fate of the infalling
material could be influenced by the amount of tangential velocity relative
to the radial one. This can be shown writing the equation of motion of a
spherically symmetric mass distribution with density $n(r)$ (\cite{peeb93}): 
\begin{equation}
\frac \partial {\partial t}n\langle v_r\rangle +\frac \partial {\partial
r}n\langle v_r^2\rangle +\left( 2\langle v_r^2\rangle -\langle v_\vartheta
^2\rangle \right) \frac nr+n(r)\frac \partial {\partial t}\langle v_r\rangle
=0  \label{eq:peeb}
\end{equation}
where $\langle v_r\rangle $ and $\langle v_\vartheta \rangle $ are,
respectively, the mean radial and tangential streaming velocity. Eq. (\ref
{eq:peeb}) shows that high tangential velocity dispersion $(\langle
v_\vartheta ^2\rangle \geq 2\langle v_r^2\rangle )$ may alter the infall
pattern. The expected delay in the collapse of a perturbation, due to
non-radial motions, dynamical friction and also taking account of a non-zero cosmological constant, may be
calculated solving the equation for the radial acceleration (\cite{kash}; \cite{kashlinsky}; \cite{kashlinsky1}; AC; \cite{peeb93}): 
\begin{equation}
\frac{dv_r}{dt}=\frac{L^2(r,\nu )}{M^2r^3}-g(r) -\eta \frac{dr}{dt}+ \frac{\Lambda}{3}r \label{eq:coll}
\end{equation}
where $L(r,\nu )$ is the angular momentum, $g(r)$ the acceleration, and $\Lambda$ the cosmological constant.
\begin{figure}
\centerline{\hbox{
\psfig{file=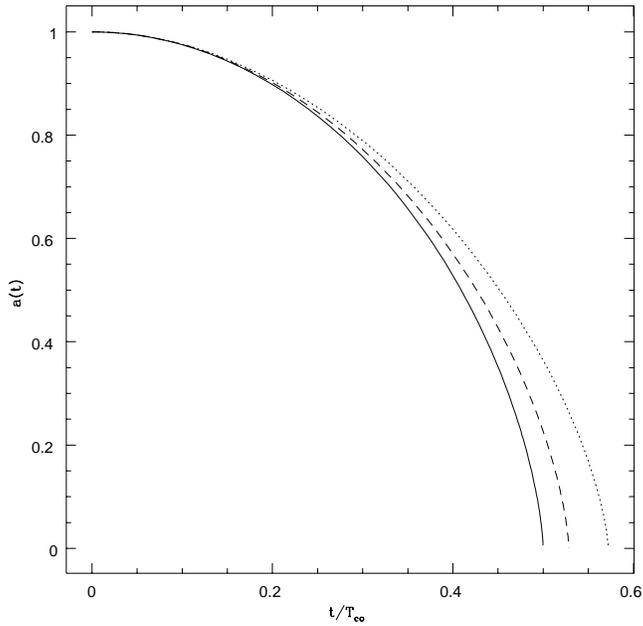,width=9cm}  
}}
\caption[]{The time evolution of the expansion parameter. The solid 
line is a(t) for the SM; the dashed line is a(t) taking account only
dynamical friction; the dotted line is a(t) taking account of the cumulative
effect of non-radial motions and dynamical friction in the case of a $\nu=2$ peak.}
\end{figure}
Writing the proper radius of a shell in terms of the expansion parameter, $%
a(r_i,t)$: 
\begin{equation}
r(r_i,t)=r_ia(r_i,t)
\end{equation}
remembering that 
\begin{equation}
M=\frac{4\pi }3\rho _b(r_i,t)a^3(r_i,t)r_i^3
\end{equation}
and that $\rho _b=\frac{3H_0^2}{8\pi G}$, where $H_0$ is the Hubble constant
and assuming that no shell crossing occurs so that the total mass inside
each shell remains constant, that is:
\begin{equation}
\rho (r_i,t)=\frac{\rho _i(r_i,t)}{a^3(r_i,t)}
\end{equation}
the Eq. (\ref{eq:coll}) may be written as: 
\begin{equation}
\frac{d^2a}{dt^2}=-\frac{H^2(1+\overline{\delta })}{2a^2}+\frac{4G^2L^2}{%
H^4(1+\overline{\delta })^2r_i^{10}a^3} -\eta \frac{da}{dt}+ \frac{\Lambda}{3}a 
\label{eq:sec}
\end{equation}
\begin{figure}
\centerline{\hbox{
2a
\psfig{file=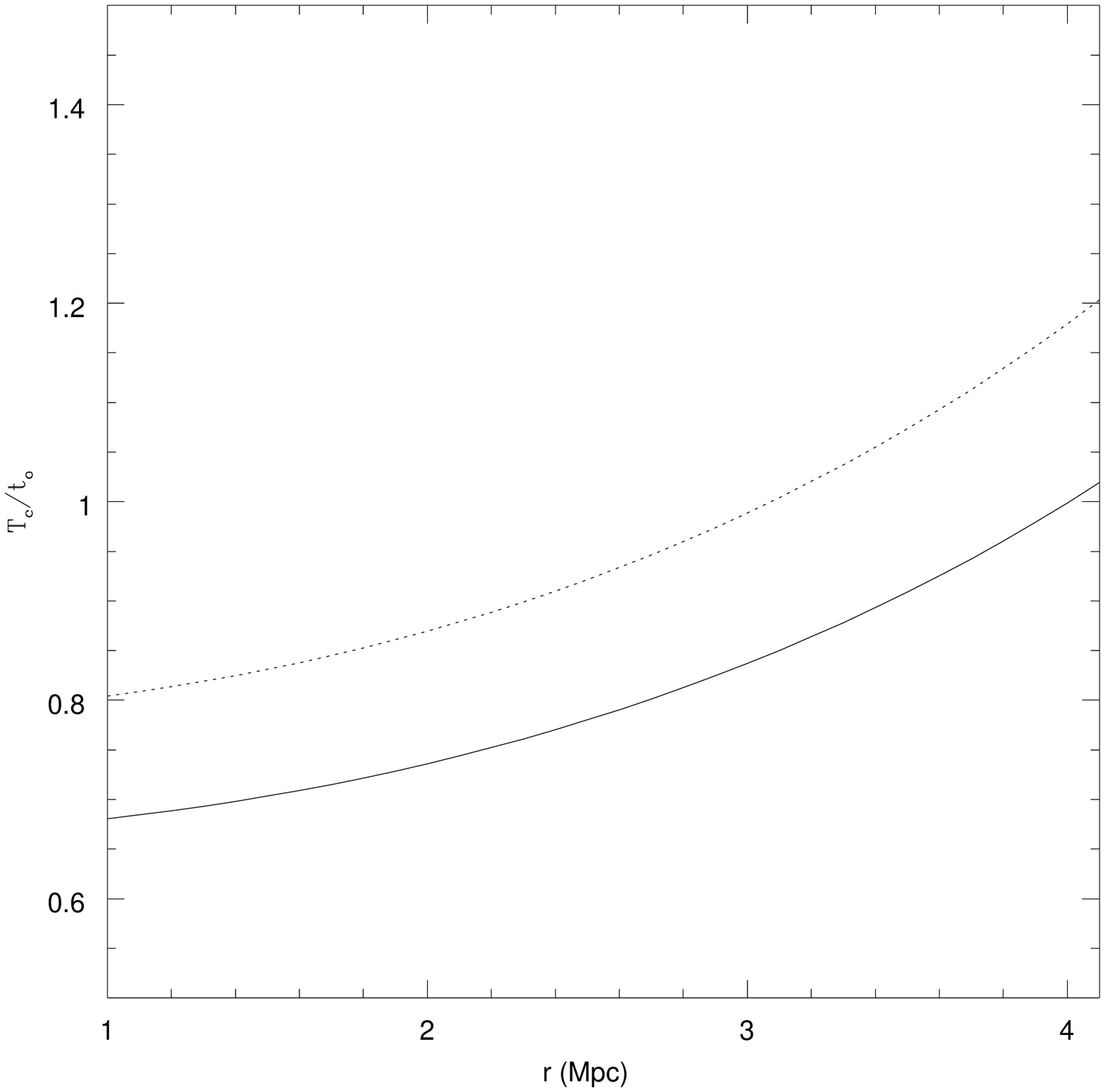,width=9cm}  
}}
\centerline{\hbox{
2b
\psfig{file=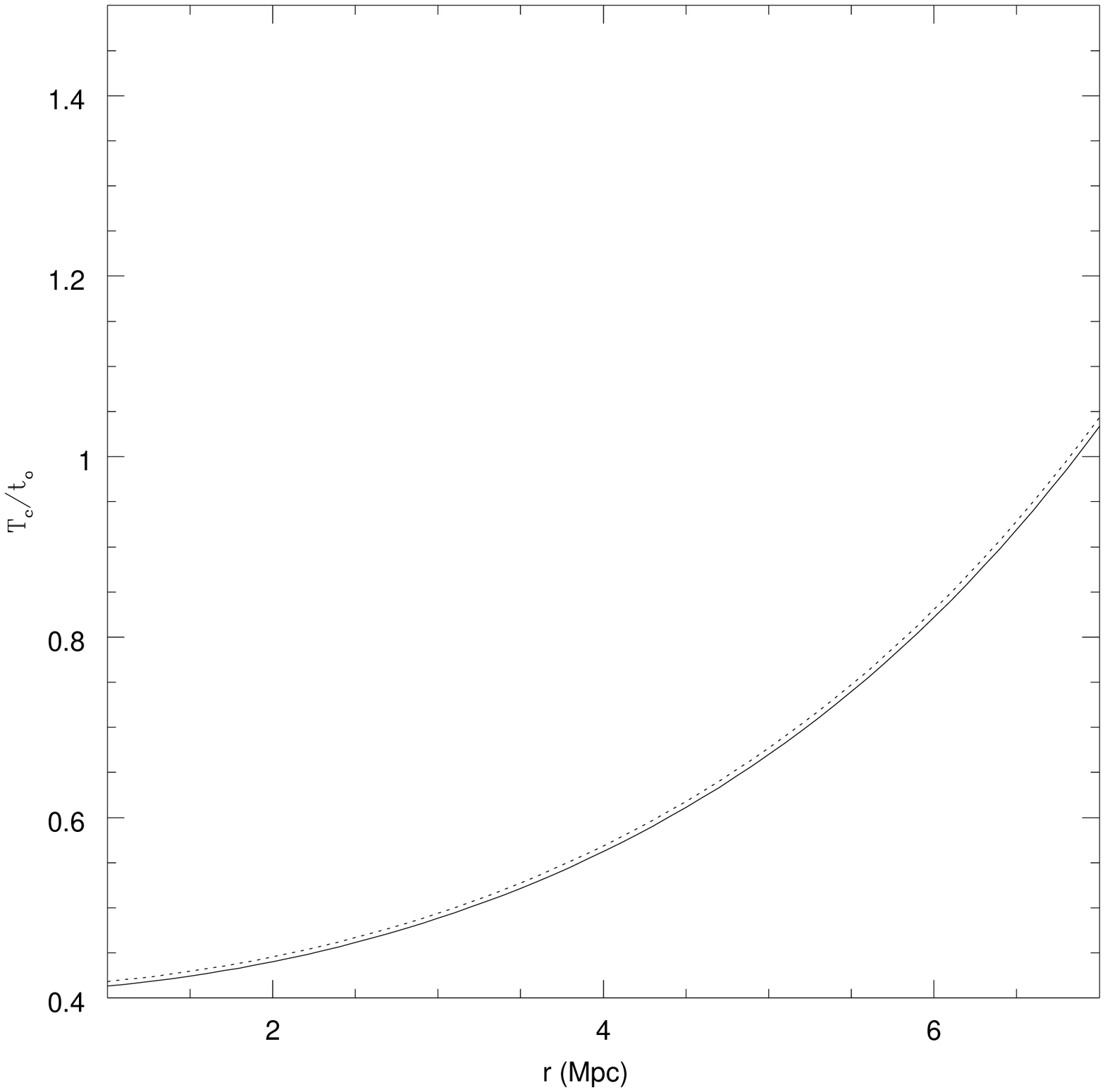,width=9cm} 
}}
\caption[]{ Fig. 2a. The time of collapse of a shell of matter in units of the age of the
universe $t_{o}$ for $\nu=2$ (dotted line) compared with SM
(solid line).
Fig. 2b. The time of collapse of a shell of matter in units of the age of the
universe $t_{o}$ for $\nu=4$ (dotted line) compared with SM (solid line).
}
\end{figure}

The equation (\ref{eq:sec}) 
may be
solved using the initial conditions: $(\frac{da}{dt})=0$, $a=a_{max}\simeq 1/%
\overline{\delta }$ and using the function $h(r,\nu )=L(r,\nu )/M_{sh}$
found in Eq. (\ref{eq:ang}), 
to obtain a(t) and the time of collapse, $T_c(r,\nu )$. As shown by \cite{gun}, this last quantity in the case of a pure 
SM (namely when tidal torques, dynamical friction and cosmological constant are not taken into account)
is given by:
\begin{equation}
T_{c0}(r,\nu )=\frac{\pi}{H_i [\overline{\delta }(r,\nu )]^{3/2}}
\footnote{As we told in the introduction, the inclusion of the peculiar velocity field changes the collapse as:
$H_i T_{c0} \simeq \frac{\pi}{(c \delta_i -\epsilon_i)^{3/2}}$
where $c$ and $\epsilon_i$ are defined in Bartelmann et al. (1993). For $\epsilon_i=0$, the collapse is shortened by a factor of $(3/5)^{3/2}$.
}
\label{eq:beef}
\end{equation} 
In Fig. 1, I show the effects of non-radial motions and dynamical friction
separately, in the case of a $\nu=2$ peak. As displayed non-radial motions have a larger effect on the
collapse delay with respect to dynamical friction.
In Fig. 2, I compare the results for the time of collapse, $T_c$, for $\nu
=2$, 4 with the time of collapse of the classical SM (Eq. \ref{eq:beef}).
As shown the presence of non-radial motions produces an increase in the time
of collapse of a spherical shell. The collapse delay is larger for low value
of $\nu $ and becomes negligible for $\nu \geq 3$. This result is in
agreement with the angular momentum-density anticorrelation effect: density
peaks having low value of $\nu $ acquire a larger angular momentum than high 
$\nu $ peaks and consequently the collapse is more delayed with respect to
high $\nu $ peaks. 

Given $T_c(r,\nu )$, I also calculated the total mass
gravitationally bound to the final non-linear configuration. 
There are at least two criteria to establish the bound region to a perturbation $\delta
(r)$: a statistical one (\cite{ryd1}), and a dynamical one (\cite{hofsha}, summarized in the following. 

In biased galaxy formation theory structures form around the 
local maxima of the density field. Every density peak binds 
a mass $M$ that can be calculated when we know the binding radius 
of the density peak. The radius of the bound region 
for a chosen density profile $\overline{\delta}( r)$ may be 
obtained in several ways. A first criterion is statistic. The binding 
radius of the region, $ r_{b}$, is given by the solution of the equation:  
\begin{equation}
< \overline{\delta} (r)> = 
< ( \overline{\delta} -<\overline{\delta}>)^{2}>^{1/2}
\end{equation}
(\cite{ryd}; \cite{ryd1}). At radius $ r << r_{b}$ the motion of particles is 
predominant toward the peak while when $r >> r_{b} $ the particle 
is not bound to the peak. Another criterion 
that can be used is dynamical. It supposes that the binding radius 
is given by the condition that a shell collapse in a time, $ T_{c}$, 
smaller than the age of the universe $ t_{0}$: 
\begin{equation}
T_{c} (r) \leq t_{0} \label{eq:temp}
\end{equation} 
(\cite{hofma}). 
This last criterion, differently from 
the previous one, contains some prescriptions particularly connected with the 
physics of the collapse process of a shell. For this reason 
I used it to calculate the binding radius. The time of collapse, $ T_{c}(r)$, 
at radius $ r$ can be obtained solving numerically Eq. (\ref{eq:sec})
for different values of $ \overline{\delta_{i}}$, the initial overdensity, 
from a given density profile $ \overline{\delta}(r)$. I use the average density 
profile given by \cite{bar}: 
\begin{equation}
\delta(r)= A \left\{ \frac{\nu \xi(r)}{ \xi(0)^{1/2}}- 
\frac{\theta( \nu \gamma, \gamma)}{\gamma \xi(0)^{1/2}(1-\gamma^{2})}
\left[\gamma^{2} \xi(r) +\frac{ R_{\ast}^{2} 
\bigtriangledown^{2} \xi(r)}{3}\right]\right\} \label{eq:dens}
\end{equation} 
where A is a constant given by the normalization of the 
perturbation spectrum, $ P(k)$,  $ \nu = \frac{\delta_{\rm c0}}{\sigma(M)} $, where $\delta_{\rm co}=1.686$ is the critical threshold for a SM, 
$\sigma(M)$ is the r.m.s. density fluctuation on the mass scale 
$M$, $ \xi(r)$ is the correlation function of two points, $ \gamma$ and $ R_{\ast}$ 
two constants obtainable from the spectrum (see \cite{bar}) and 
finally $ \theta ( \gamma \nu, \gamma)$ is a function given in 
the quoted paper (eq. 6.14). Given the average density 
profile the average density inside the radius $ r$ in  a spherical 
perturbation is given by Eq. (\ref{eq:denss}). 

Finally, I calculated the binding radius, $r_{b}(\nu )$, for a SM, calculating $T_{\rm co}(r)$ by means of   
Eq.~(\ref{eq:beef}) 
and the density profile given in Eq.~(\ref{eq:dens}) and then applying the condition $T_{\rm co}(r) \leq t_o$.
I repeated the calculation for 
$ 1.7 <\nu< 4 $.
Then I repeated the calculation using $T_c(r)$, the collapse time that takes into account non-radial motions and dynamical friction.
I found a relation between $\nu $ and the mass of the cluster using the
equation: $M=\frac{4\pi }3r_b^3\rho _b$.
The result is the plot in Fig. ~3 for the binding radius $ r_{b} $ versus $ \nu$.

In fig. 4, I compare the peak
mass obtained from SM, using \cite{hofsha} criterion,
with that obtained from the model taking into account non-radial motions, dynamical friction and $\Lambda \neq 0$. As
shown for high values of $\nu $ ($\nu \geq 3$) the two models give the same
result for the mass while for $\nu <3$ the effect of non-radial motions
produces less bound mass with respect to SM. 
Decreasing the effect of non-radial
motions produces a decrease in the bound mass.

The situation represented in the previous three figures may be summarized as follows: dynamical friction and non-radial motions delays the collapse of 
perturbations. Both effects act in the direction of delaying structure collapse, so that their effects add.
The effects have a similar magnitude, but non-radial motions
induce a slight larger delay in collapse. As a consequence of this delay of collapse the matter bound to structures is less than what expected in the case of SM.

\begin{flushleft}
{\it 1.9.5 The threshold of collapse $\delta_{\rm c}$}
\end{flushleft}


In this section, I am going to show how dynamical friction and tidal fields influence 
the critical overdensity threshold for the
collapse, $\delta _c$, which is not constant as in a SM
but it depends on mass.
An analytic determination of $\delta _c(\nu )$ can be obtained following a
technique similar to that used by \cite{bart}. 

\begin{figure}[ht]
\psfig{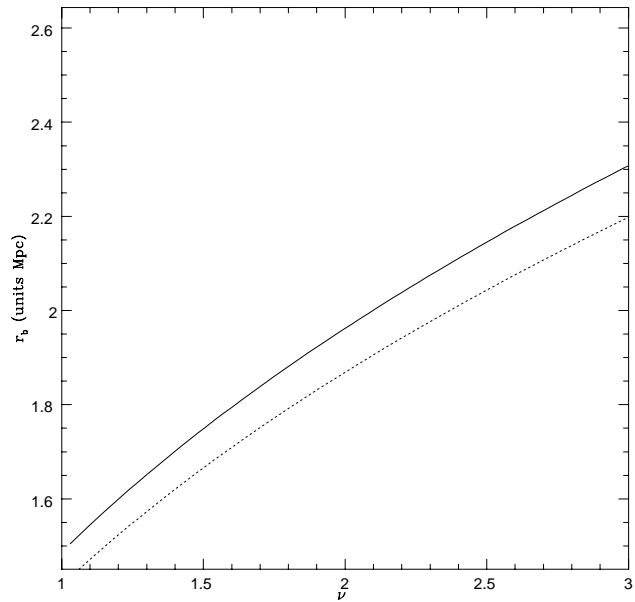}
\caption[h]{Variation of the binding radius $ r_{b} $ with $ \nu $.
The solid line is the binding radius in the SM,
while the dashed line is the same as in presence of non-radial motions dynamical friction and $\Lambda \neq 0$, for $\nu=2$.
}
\end{figure}

Using Eq.~(\ref{eq:sec}) it is possible to obtain the value of the expansion parameter of
the turn around epoch, $a_{max}$, which is characterized by the condition $\frac{da}{dt}=0$. Using the relation between $v$ and $\delta _i$, in linear
theory (\cite{pee1}), I find:

\begin{eqnarray}
B(M)&=&\delta _{\rm c}=\delta _{\rm co}
\nonumber \\
& &
\left[ 1+
\int_{r_{\rm i}}^{r_{\rm ta}}  \frac{r_{\rm ta} L^2 \cdot {\rm d}r}{G M^3 r^3}+\frac{\lambda_{o}}{1-\mu(\delta)}+\Lambda \frac{r_{\rm ta} r^2}{6 G M}
\right] 
\label{eq:barriera} 
\end{eqnarray}
%
%
where $\delta _{co}=1.686$ is the critical threshold for SM, $r_{\rm i}$ is the initial radius, $r_{\rm ta}$ is the turn-around radius, 
$\lambda_{o}=\epsilon_o T_{co}$ and $ \mu(\delta)$ is given in \cite{colande} (Eq. 29). The quantity $L$ appearing in Eq. ~(\ref{eq:barriera}) is 
the total angular momentum acquired by the proto-structure during evolution that is calculated as shown in section 1.9.3. 
%
%
%
%
The result of the calculation is shown
in Fig. 5, where I plot $\delta_{\rm c}(\nu)$ obtained by
means of the model of the present paper together with that obtained by \cite{shto} (ST) using
an ellipsoidal collapse model. The dashed line
represents $\delta_{\rm c}(\nu)$ obtained with the present model, while the
solid line that of ST. Both models show that the threshold
for collapse
decreases with mass and when $\nu $ $>3$ the
threshold assume the typical value of the SM.
In other words, this means that, in order to form
structure, more massive peaks must
cross a lower threshold,
$\delta_c(\nu)$, with respect to under-dense ones.
At the same time, since the
probability to find high peaks is larger in more dense regions, 
this means that, statistically, in order to form structure, 
peaks in more dense
regions may have a lower value of the threshold, $\delta_c(\nu)$, with respect
to those of under-dense regions.
This is due to
the fact that less massive objects are more influenced by external tides, and
consequently they must be more overdense to collapse by a given time.
In fact,
the angular momentum acquired by a shell centered on a peak
in the CDM density distribution is anti-correlated with density: high-density
peaks acquire less angular momentum than low-density peaks
(\cite{hofma}; \cite{ryd}; \cite{ryd1}).
A larger amount of
angular momentum acquired by low-density peaks
(with respect to the high-density ones)
implies that these peaks can more easily resist
gravitational collapse and consequently it is more difficult for them to form
structure.
This is in agreement with  
\cite{audit}, \cite{peebles90}, which pointed out that the
gravitational collapse 
is slowed down by the  effect of the shear
rather than fastened by it (as sustained by other authors).
Therefore, on small scales, where the shear is statistically greater,
structures need, on average, a higher  density contrast to collapse. This results in a tendency for less dense
regions to accrete less mass, with respect to a classical SM,
inducing a {\it biasing} of over-dense regions towards higher mass.

\begin{figure}
\psfig{file=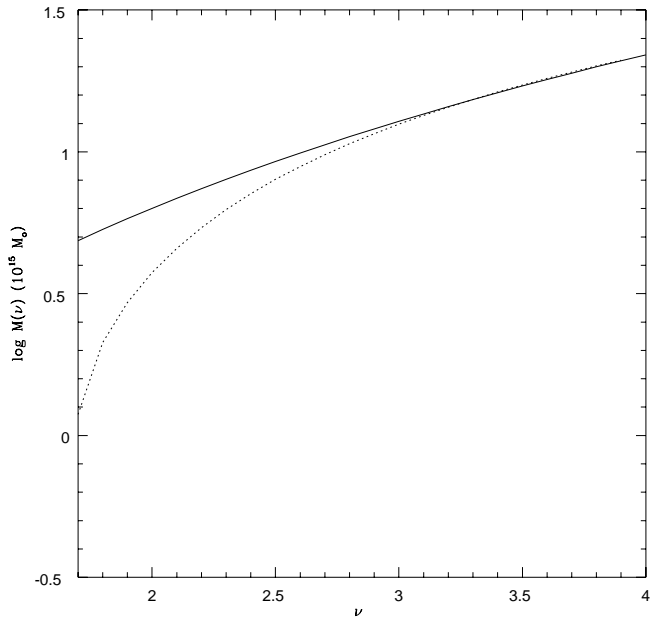,width=9cm,height=9cm}
\caption[]{The mass accreted by a collapsed perturbation, in units of
$10^{15}M_{\odot}$, taking into account non-radial motions, dynamical friction and a non-zero cosmological constant (dotted line)  
compared to SM mass (solid line).}
\end{figure}

\begin{figure}[ht]
\psfig{file=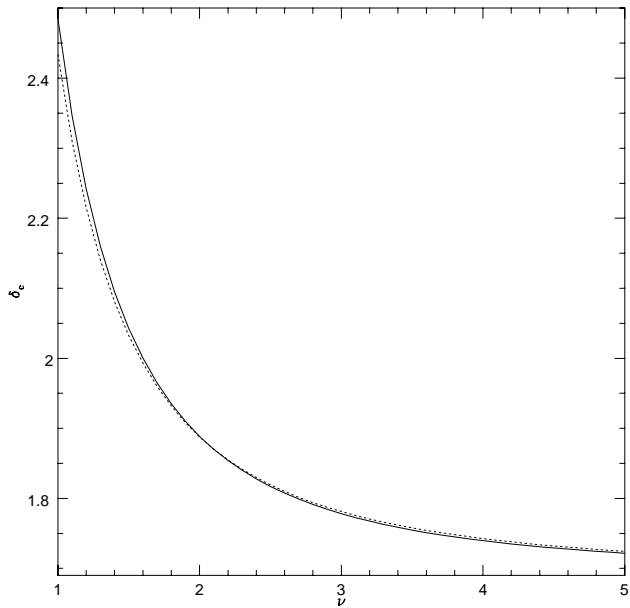,width=9cm,height=9cm}
\caption[]{The critical threshold, $\delta_{c}(\nu)$ versus $\nu$. The dashed line is obtained with the model of the present paper while the solid line is that of ST.} 
\end{figure}

\begin{flushleft}
{\it 1.10 Mass function}
\end{flushleft}


One of the most important quantities in cosmology is the mass function or multiplicity function. It can be described by the relation:
\begin{equation}
dN = n( M ) dM
\end{equation}
that is the number of objects per unit volume, having a mass in the range $ M $ ed $ M + dM $. The multiplicity function can also be used to define the luminosity function after having fixed the ratio $ \frac{ M}{ L} $.
Obtaining the mass function starting from that of luminosity is complicated since the ratio $ \frac{ M }{ L } $
is known with noteworthy uncertainty and it is different for different objects and moreover the luminosity function for objects like galaxies depend on the morphological type.  
Finally trying to determine the luminosity function observatively is problematic 
(see for example \cite{efs2}).\\
For the above reasons, the theoretical determination of the mass function is very important. One of the most successful study in the subject is that of \cite{pre}. This theory is based upon these hypotheses:
\begin{itemize}
\item
The linear density field is described by a stochastic Gaussian field. The statistics of the matter distribution is Gaussian.
\item 
The evolution of density perturbations is that described by the linear theory. Structures form in those regions where the overdensity linearly evolved and filtered with a top-hat filter exceeds a threshold $ \delta_{c} $ ($ \delta_{c} =1.68 $, obtained from the spherical collapse model (\cite{gun})). 
\item 
for $ \delta \geq \delta_{c} $ regions collapse to points. The probability that an object forms at a certain point is proportional to the probability that the point is in a region with $ \delta \geq \delta_{c} $ given by:
\begin{equation}
P ( \delta , \delta_{c} )=
\int_{\delta_{c}}^{\infty}d \delta \frac{1}{\sigma
(2 \pi)^{\frac{1}{2}}}
 exp \left(-\frac{\delta^{2}}{2 \sigma^{2}}\right)
\end{equation}
The multiplicity function is given by:
\begin{equation}
\rho ( M, z )= - \rho_{0}\frac{ \partial P}{\partial M } dM =
n( M ) M dM \label{eq:ps}
\end{equation}
\end{itemize}
If we add the conditions
$ \Omega=1 $, $ \left|\delta_{k}\right|^{2} \propto k^{n}$,
the Press-Schechter solution is autosimilar and has the form:
\begin{eqnarray}
\rho ( M, z) =\frac{ \rho}{\sqrt{2 \pi}}
\left(\frac{n+3}{3}\right) \left( \frac{M}{M_{\ast }}( z)
\right)^{\frac{n+3}{6}} \nonumber \\
\times exp \left[ -\frac{1}{2} \frac{M}{M_{\ast}}( z )^{
\frac{n+3}{3}}\right]\frac{dM }{M}
\end{eqnarray}
where $ M_{\ast}(z ) \propto (1+z)^{-\frac{6}{n+3}} $. 
Several are the problems of the theory: 
\begin{itemize}
\item 
{\bf Statistical problems:} in the limit of vanishing smoothing radii,
or of infinite variance, the fraction of collapsed mass, 
asymptotes to 1/2. This is a signature of
linear theory: only initially overdense regions, which constitute half
of the mass, are able to collapse. Nonetheless, underdense regions can
be included in larger overdense ones, or, more generally,
non-collapsed regions have a finite probability of being included in
larger collapsed ones; this is commonly called {\it cloud-in-cloud
problem}.  PS argued that the missing mass would accrete on the formed
structures, doubling their mass without changing the shape of the MF;
however, they did not give a true demonstration of that. Then, they
multiplied their MF by a ``fudge factor'' 2. Other authors 
used to multiply the MF by a factor $(1+f)$, with $f$
denoting the fraction of mass accreted by the already formed
structures.
\item
{\bf Dynamical problems:} the heuristic derivation of the PS MF
bypasses all the complications related to the highly non-linear
dynamics of gravitational collapse. Spherical
collapse helps in determining the $\delta_c$ parameter and in identifying
collapsed structures with virialized halos. However, the PS procedure
completely ignores important dynamical elements, such as the role of
tides and the transient filamentary geometry of collapsed structures.
Moreover, supposing that every structure virializes just after
collapse is a crude simplification: when a region collapses, all its
substructure is supposed by PS to be erased at once, while in
realistic cases the erasure of substructures is connected to the
two-body interaction of already collapsed clumps, an important piece
of gravitational dynamics which is completely missed by the PS
procedure.
\item
{\bf Geometrical problems:} to estimate the mass function from
the fraction of collapsed mass at a given scale it is
necessary to relate the mass of the formed structure to the resolution

In practice,
the true geometry of the collapsed regions in the Lagrangian space
(i.e. as mapped in the initial configuration) can be quite complex,
especially at intermediate and small masses; in this case a different
and more sophisticate mass assignment ought to be developed, so that
geometry is taken into account. For instance, if structures are
supposed to form in the peaks of the initial field, a different and
more geometrical way to count collapsed structures could be based on
peak abundances.
\end{itemize}

Despite all of its problems, the PS procedure proved successful, as
compared to N-body simulations, and a good starting point for all the
subsequent works on the subject (\cite{efs3}; \cite{efs4}, 
Bond et al. (1991), 
\cite{whi1},
\cite{jain}, 
\cite{lac}, \cite{efs5}, 
\cite{bond2}.  Most authors reported the PS
formula to fit well their N-body results; nonetheless, all the authors
agree in stating the validity of the PS formula to be only
statistical, i.e.  the existence of the single halos is not well
predicted by the linear overdensity criterion of PS (see in particular
Bond et al. 1991)). 
There are however some exceptions to this general agreement: \cite{brainar} reported their MF, based on a CDM spectrum, to
be very similar to a power-law with slope $-2$, different from the PS
formula both at small and at large masses.  Several authors, \cite{jain}, \cite{gel} and \cite{ma}
noted that, to make the PS formula agree with their simulations (based
on CDM or CHDM spectra), it is necessary to lower the value of the
$\delta_c$ parameter as redshift increases.  The same thing was found by
\cite{kly1}, but was interpreted as an artifact of their
clump-finding algorithm. Recent simulations seem to confirm this trend.

Lacey \& Cole (\cite{lac1}) extended the comparison to N-body simulations to
the predictions for merging histories of dark-matter halos; they found again a good agreement between theory and
simulations. This fact is noteworthy, as merging histories contain
much more detailed information about hierarchical collapse.
Several improvements of the theory exists: \cite{lac} (the so called Extended Press \& Schechter (EPS) formalism; \cite{shto1}). 

\begin{flushleft}
{\it 1.11 CDM, HDM and others cosmogonies}
\end{flushleft}
 

The study of origin and formation of structure in the universe has been historically fundamentally framed into two theories: the CDM theory, in which WIMPS constitutes the main part of Dark Matter, and HDM in which neutrinos dominate. As we are going to see, structure formation in these scenarios is completely different since WIMPS and neutrinos are subject to different physical phenomena and then the transfer function is noteworthy different in the two cases. 
Both theories have the same starting points:\\
1) The universe is fundamentally constituted by Dark Matter (WIMPS in the CDM model and neutrinos
in the HDM) and $\rho =\rho_{c} $ (  $\Omega =1$). \\
2) Baryons give a small contribution to the mass of the universe.\\
3) Fluctuations originating in the primordial universe are adiabatic, scale invariant, $n=1$, and Gaussian (\cite{efs}).\\
If universe is dominated by neutrinos with mass $ m_{\nu} = 30 ev $ the transfer function is determined by the free-streaming (or Landau damping) of neutrinos. This phenomenon consists in the smoothing of inhomogeneities in the primordial universe (due to perturbations in the acollisional components) because of the motion of neutrinos from overdense to underdense regions. Neutrinos diffusion and the smoothing of inhomogeneities is possible only before $ t = t_{eq} $.
After this epoch, there is no longer free-streaming but the density perturbation has definitely changed by its previous action. Free-streaming scale or mass can be estimated calculating the distance covered from a particle decoupled from plasma. Results that one obtains for the free-streaming scale and mass is (\cite{kol}): 
\begin{equation}
\lambda_{FS} \approx 40 Mpc (m_{\nu}/30 ev )^{-1} \hspace{1cm}
M_{FS} \approx 10^{15}(m_{\nu}/ 30 ev)^{-2} M_{0}
\end{equation}
Because of free-streaming of neutrinos the HDM spectrum is characterized by a cut-off at short wave-length. The result is that the first objects that form are superclusters and structure formation proceeds because of fragmentation. 
\cite{zel} showed that the first structure to form are flat and were called "pancakes". After formation, these objects enter in the non-linear phase along one of the axes and baryons inside start to collide and dissipate their gravitational energy. Galaxies form for fragmentation processes. Structure formation follows a 'top-down' scheme, that means that larger objects (e.g., clusters) form before, while smaller objects (e.g., galaxies) later. N-body simulations of HDM universes 
(\cite{fre}; \cite{cent} showed that on scales larger than 10 Mpc structures is qualitatively similar to voids and to the filamentary structure that is visible in the CFA, but the clustering measured in N-body simulations is larger than that observed in the CFA.
When one tries to reproduce the observed correlation function, one arrives to the conclusion that pancaking should have happened at redshift $ z \leq 1 $, in disagreement with observed galaxies having $ z \geq 1 $ and QSO with $ z \geq 3 $.
A further problem of the model is that of the peculiar velocity that are smaller than values obtained from observations (\cite{kol}). \\
The HDM model after a series of studies in the '80s has been abandoned for the problems it has and replaced by another model, the CDM, which is in better agreement with observations. The CDM model has a spectrum without a cut-off at short wave -length (at least till scales much smaller than galactic scales) because the damping scale is unimportant for WIMPS with mass $ > $ 1 Gev. Structure formation is typically hyerachical: from smaller scale structure to larger ones. This scheme is  a 'bottom-up' scheme. When the CDM model was introduced it obtained noteworthy successes in the description of the characteristic of the universe (clustering statistics of galaxies, peculiar velocities, CMBR fluctuations) from the galactic scale on (\cite{pee2}; \cite{blu}; \cite{bar};
\cite{whi}; \cite{fre1}; \cite{efs}). The model has shown some weak points, when compared with more and more precise data.\\
The reason of the success of the CDM model is fundamentally due to the fact that WIMPS interact with matter by means of gravity only, and does not feel the effect of pressure forces due to interaction with radiation (to which matter components are subject). Structure formation starts before in the CDM component, at $ t< t_{eq} $, which give rise to the potential wells in which baryonic matter can then fall. It is important to notice that, in order to reproduce observations, an additive hypothesis must be added: the biasing hypothesis, that can be summarized in: {\it light does not trace mass}. \footnote{If this hypothesis is not introduced the discrepancies between data and theory can be reduced assuming a value of $ H_{0} \approx 25 km/Mpc s $}. Typical problems of the CDM model in absence of biasing are the too high values for the r.m.s. of peculiar velocity of couple of galaxies (values of 1000 km/s vs. $ 300 \pm 50 $ km/s observed).
Another problem is the correlation length, $r_{0} $, in N-body simulations, for the correlation function $ \xi $ which is equal to $ 1.3 h^{-1} Mpc $, smaller than the observed value $ 5.5 h^{-1} Mpc $.
On the other side, if, in order to eliminate the quoted problems one introduce the biasing hypothesis, there is the supplementary problem of finding a physical mechanism that explains the origin of bias. Several conjectures has been proposed (\cite{ree}; \cite{sil1}; \cite{dek}) but there is no full agreement on them. 

Summarizing, one can tell that
although at the beginning the standard form of CDM was very successful in describing the structures observed in the
Universe (galaxy clustering statistics, structure formation epochs, peculiar velocity flows) (\cite{pee2}; \cite{blu}; \cite{bar}; \cite{whi}; \cite{fre1}; \cite{efs}) recent measurements
have shown several deficiencies in the model, at least, when any bias of the distribution of galaxies relative to the mass is constant with scale (see \cite{bab}; \cite{bower}; \cite{del}, \cite{dell}). Some
of the most difficult problems that must be reconciled  with the theory are:
\begin{itemize}
\item the magnitude of the
dipole of the angular distribution of optically selected galaxies 
(\cite{kai});
\item the possible observations of clusters of
galaxies with high velocity dispersion at $z\geq 0.5$ (\cite{evrar});
\item the
strong clustering of rich clusters of galaxies, $ \xi_{cc}(r) \simeq (r / 25h^{-1}Mpc)^{-2}$, 
far in excess of CDM predictions (\cite{bah1});
\item the X-ray temperature distribution function of clusters,
over-producing the observed clusters abundances (\cite{bart}); 
\item the
conflict between the normalization of the spectrum of the perturbation which is required by different types of observations;
\item 
the incorrect scale dependence of the galaxy correlation
function, $\xi (r)$, on scales $10$ to $100$ $h^{-1} Mpc$, having $\xi (r)$ too little power on the large scales compared to the power on smaller scales 
(\cite{maddox}; \cite{san1}; 
\cite{pea}; \cite{pea1}).
\item
Normalization obtained from COBE data (\cite{smo}) on scales of the order of $10^3Mpc$ 
requires $\sigma_8=0.95\pm 0.2$, where $\sigma_8$ is the rms value of $\frac{\delta M}{M}$ in a sphere of $8 h^{-1}$Mpc.
Normalization on scales $10$ to $50Mpc$ obtained from QDOT and POTENT (\cite{dek1}) requires that $\sigma _8$ is in the range $0.7\div 1.1$, which is compatible with COBE normalization
while the observations of the pairwise velocity dispersion of galaxies on
scales $r\leq 3Mpc$ seem to require $\sigma _8<0.5$. 
\item 
Another problem of CDM
model is the incorrect scale dependence of the galaxy correlation
function, $\xi (r)$, on scales $10$ to $100$ $Mpc$, having $\xi (r)$ too
little power on the large scales compared to the power on smaller scales.
The APM survey (\cite{maddox}), giving the galaxy angular
correlation function, the 1.2 Jy IRAS power spectrum, the QDOT survey
(\cite{san1}), X-ray observations (\cite{lah1}) and radio
observations (\cite{pea}; \cite{pea1}) agree with the quoted
conclusion. As shown in recent studies of galaxy clustering on large scales
(\cite{maddox}; \cite{san1}) the
measured rms fluctuations within spheres of radius $20h^{-1}Mpc$ have
value 2-3 times larger than that predicted by the CDM model.\\
\item Density profiles of CDM halos: the cusp obtained from numerical 
simulations seems too steep. 
-\item simulations might yield too many satellites for galaxies like our own. 
Though this second problem may have been the result of bad comparison 
of simulations with observations. 
This yielded a surge of interest in the last 2-3 years for Warm Dark matter and 
slightly collisional matter.
\end{itemize}

These discrepancies between the theoretical predictions of the CDM model
and  the observations led many authors to conclude that the shape of the
CDM spectrum is incorrect and to search alternative models
(\cite{pee3}; \cite{sha}; \cite{vald}; 
\cite{holt}; \cite{tur}; \cite{schae}; \cite{cen}; 
\cite{bower}).\\
Alternative models with more large-scale power than CDM have been introduced
in order to solve the latter problem. Several authors have lowered the matter density under
the critical value ($ \Omega_m < 1$) ($\Omega_m \simeq 0.3$) (this model is called open cold dark matter model (OCDM)) and others (\cite{pee3};
Efstathiou et al. 1990a; \cite{tur}) have also added a
cosmological constant in order to retain 
a flat Universe ($ \Omega _m + \Omega _\Lambda = 1$).
This model is known as $\Lambda$CDM model.
The spectrum of the matter density is specified by the
transfer function, but its shape is
affected because of the fact that the epoch of
matter-radiation equality is earlier,
$1+z_{eq}$ being increased by a factor $1/\Omega_{m}$. 
The epoch of
matter-radiation equality is earlier, because $1+z_{eq}$ is increased by a
factor $1/\Omega _m$.
Around the epoch $z_\Lambda $ the growth of the
density contrast slows down and ceases after $z_\Lambda $.
As a consequence the
normalization of the transfer function begins to fall, even if its shape
is retained (and pushes its imprint to larger scales). 
Mixed Dark Matter models (MDM) (\cite{sha}; \cite{vald}; 
\cite{holt}; \cite{schae})
increase
the large-scale power because neutrinos free-streaming damps the power on
small scales. Alternatively, changing the primeval spectrum
several problems of CDM are
solved (\cite{cen}). For example, in the $\tau$CDM model the needed changes
in the power spectrum may be obtained in $\Omega=1$ CDM models if matter-radiation equivalence is delayed, such as by the 
addition of an additional relativistic particle species. 
Finally it is possible to assume
that the threshold for galaxy formation
is not spatially invariant but weakly modulated ($2\%-3\%$ on scales $%
r>10h^{-1}Mpc$) by large scale density fluctuations, with 
the result that the clustering on
large-scale is significantly increased (\cite{bower}).

\begin{flushleft}
{\it 1.12 Constraints from recent astrophysical observations}
\end{flushleft}

In the past decade we have witnessed spectacular
progress in precision measurements in astrophysics as a result
of significant improvements in terrestrial and extraterrestrial 
instrumentation. The (second phase of the) 
Hubble telescope opened up novel paths
in our quest for understanding the Universe, by allowing observations
on distant corners of the observable Universe that were not accessible
before. 

From the point of view of interest to particle physics, 
the most spectacular claims from astrophysics came 8 years ago
from the study of distant supernovae (redshifts $z \sim 1$) 
by two independent groups~\cite{supernovae}.
These observations pointed towards
a current era acceleration of our Universe, something that 
could be explained
either by a non-zero cosmological {\it constant} in a Friedman-Robertson-Walker-Einstein Universe, or in general by a non-zero {\it dark energy} component,
which could even be relaxing to zero (the data are consistent
with this possibility). This claim, if true, could revolutionize our
understanding of the basic physics governing fundamental interactions
in Nature. Indeed, only a few years ago, particle theorists were trying to 
identify (alas in vain!) an exact symmetry of nature that could 
set the cosmological constant (or more generally the vacuum energy) to zero.
Now, astrophysical observations point to the contrary.
The skeptics may question the accuracy of the supernovae 
observations, however, there is additional evidence from quite different in origin astrophysical observations, those associated with the measurement of 
the cosmic microwave background radiation (CMB), which point towards
the fact that $73$ \% of the Universe vacuum energy consists of 
a dark (unknown) energy substance, in agreement with the (preliminary) 
supernovae observations. Moreover, recently~\cite{recentsn} 
two more distant supernovae have been discovered 
($z > 1$), exhibiting  similar features as the previous measurements,
thereby supporting the geometric interpretation on the acceleration
of the Universe today, and arguing against the nuclear physics or
intergalactic dust effects. 

Above all, however, there are the very recent data 
from a new probe of Cosmic Microwave Background Radiation Anisotropy 
(Wilkinson Microwave Anisotropy Probe (WMAP))~\cite{wmap}.
In its first year of running WMAP
measured CMB anisotropies to an unprecedented accuracy 
of billionth of a Kelvin degree, thereby correcting previous measurements
by the Cosmic Background Explorer (COBE) satellite~\cite{cobe} 
by several orders of magnitude.
This new satellite experiment, therefore, opened up a new era
for astroparticle physics, given that such accuracies allow 
for a determination (using best fit models of the Universe) 
of cosmological parameters~\cite{spergel}, and 
in particular cosmological densities, 
which, as we shall discuss in this review, 
is quite relevant for constraining models of particle physics
to a significant degree. 

The WMAP satellite experiment determined the most important 
cosmological parameters that could be of relevance to 
particle physicists, namely~\cite{spergel}: the 
Hubble constant, and thus the 
age of the Universe,
the thickness of the last scattering surface, the 
dark energy and dark matter content of the Universe 
(to an unprecedented accuracy), 
confirming the earlier claims from supernovae Ia data~\cite{supernovae},
and 
provided evidence for early reionization ($z \sim 20$),
which, at least from the point of view of large scale structure 
formation, excludes Warm Dark Matter particle theory models. 

An important comment 
concerns the dark energy component ($73$ \% ) of the Universe.
The WMAP measured equation of state 
for the Universe $p = w\rho $, 
with $p$ the pressure and $\rho$ the energy density, implies 
$-1 \le w < -0.78$ (assuming the lower bound for theoretical reasons, 
otherwise the upper limit may be larger~\cite{spergel}). 
For comparison we note that $w=-1$ characterizes a perfect fluid Universe
with non-zero, positive, cosmological constant.
As we shall remark, supergravity quintessence models do have this feature
of $w \to -1$, and it may well be that by exploiting further the data on  
this dark energy component of the Universe one may arrive at the physically correct supergravity model which could constrain the supersymmetric 
particle physics models. 

The results of the WMAP analysis (alone),including directly measurable and
derived quantities,  are summarized 
in the tables appearing in figures \ref{table1spergel},\ref{table2}.

\begin{figure}
\centerline{\hbox{
\psfig{file=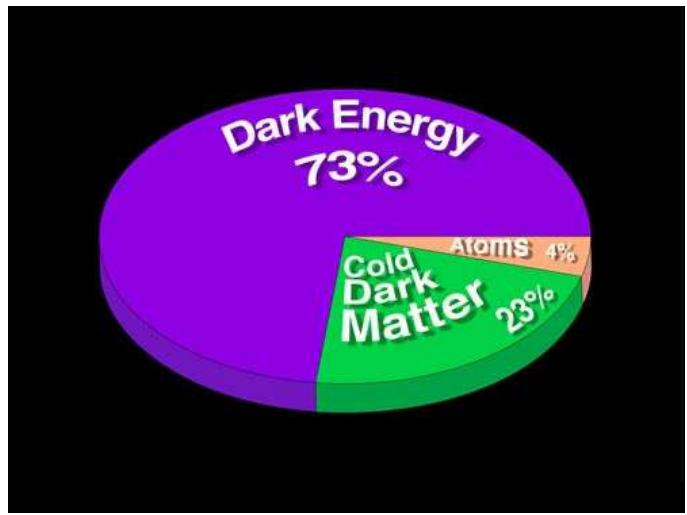,width=9cm}  
}}
\caption[]{
The energy content of our Universe as obtained
by fitting data of WMAP satellite. The chart is in 
perfect agreement with earlier claims made by direct measurements
of a current era acceleration of the Universe from distant 
supernovae type Ia (courtesy of  
http://map.gsfc.nasa.gov/).
}
\end{figure}

One therefore obtains the chart for the energy and 
matter content of our Universe depicted in figure 6.
This chart is in perfect agreement  
with 
{\it direct} evidence on acceleration of the Universe 
(and hence cosmological constant) from Supernovae Ia Data~\cite{supernovae}.
\begin{figure}
\centerline{\hbox{
\psfig{file=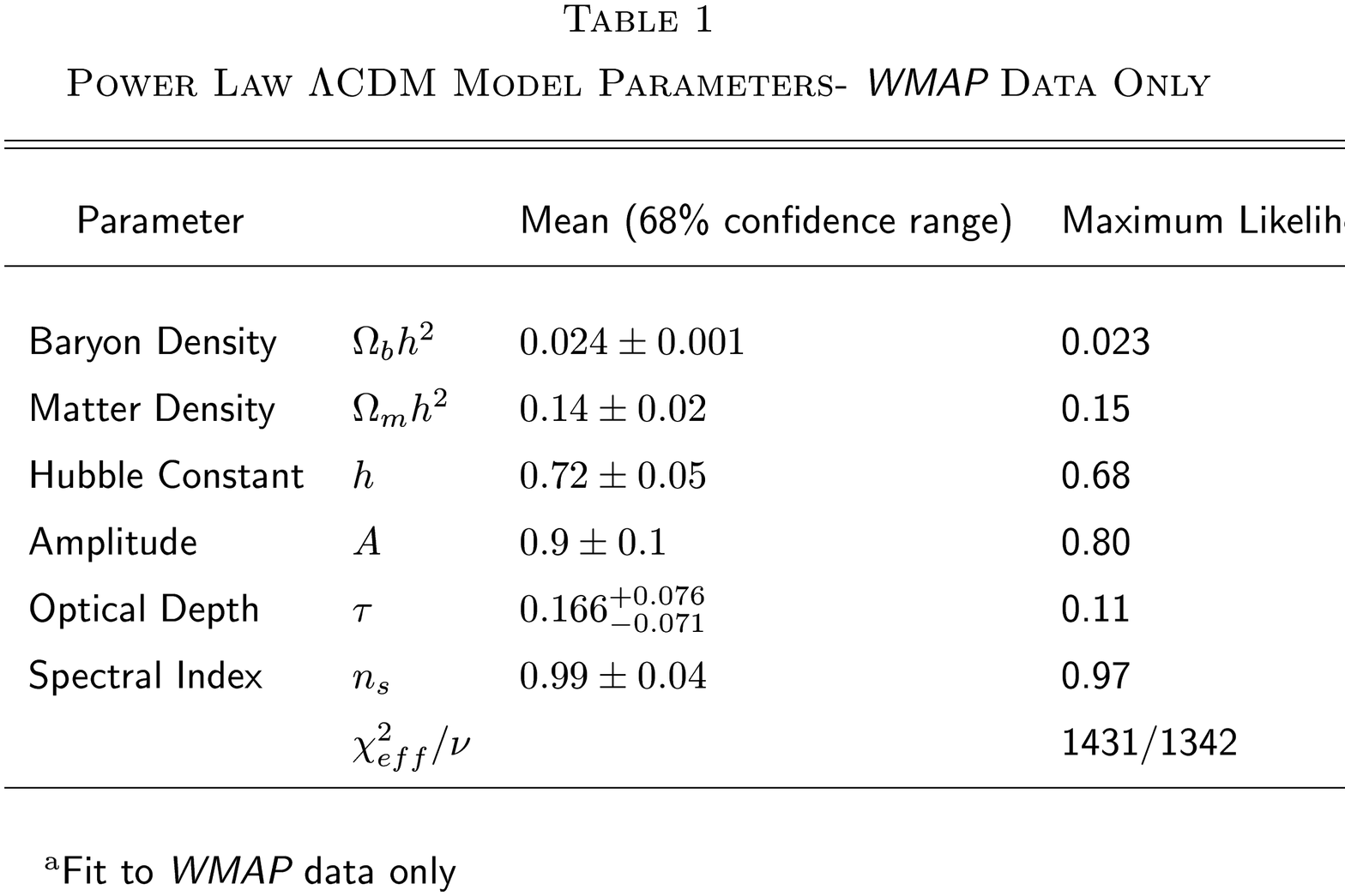,angle=-90,width=9cm}  
}}
\caption{Cosmological parameters measured by WMAP (only):directly measurable quantities
~\cite{spergel}.
}
\label{table1spergel}
\end{figure}

\begin{figure}
\centerline{\hbox{
\psfig{file=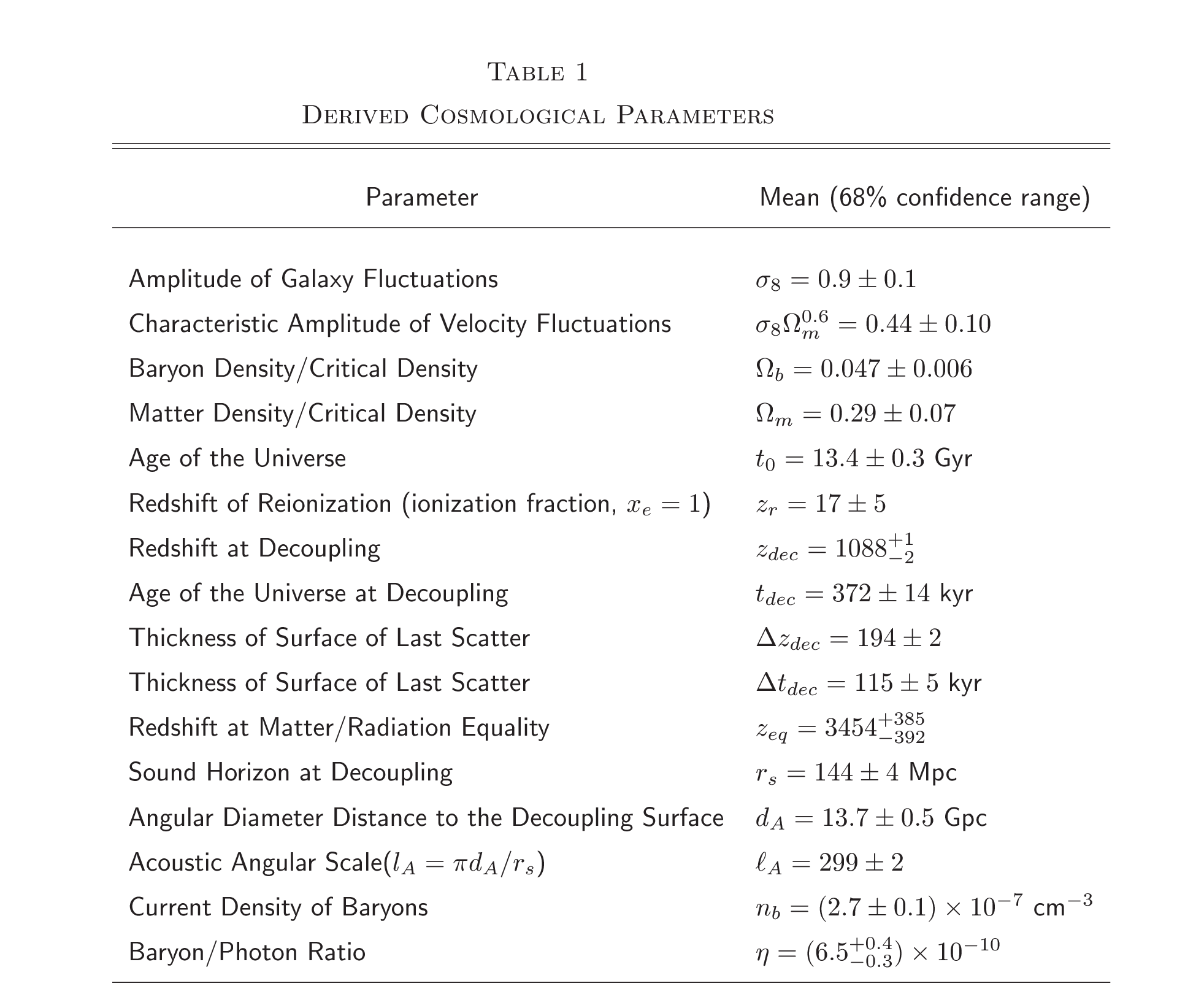,width=9cm}  
}}
\caption{Cosmological parameters measured by WMAP (only):derived 
quantities
~\cite{spergel}.
}
\label{table2}
\end{figure}

It should be stressed that 
the interpretation 
of the supernovae data 
is based on a  {\it best fit} Friedmann-Robertson-Walker (FRW) 
Universe~\cite{supernovae}: 
\begin{equation} 
0.8\Omega_M - 0.6\Omega_\Lambda \simeq -0.2 \pm 0.1~,
\quad {\rm for}~\Omega_M \le 1.5
\label{o1}
\end{equation}
with $\Omega_{M,\Lambda}$ corresponding to the matter and
cosmological matter densities. 
Assuming a flat model (k=0), 
$\Omega_{\rm total} = 1$, as supported by the 
CMB data, the SNIa data alone imply: 
\begin{equation} 
\Omega_M^{Flat} =
0.28^{+0.09}_{-0.08}~(1\sigma~stat.)^{+0.05}_{-0.04}~(identified~syst.)
\label{o2}
\end{equation}
The deceleration parameter defined as
$q \equiv - \frac{{\ddot a} a}{ {\dot a}^2}$ , where $a$ is the
cosmic scale factor, receives the following form if we omit the
contribution of photons which is very small,
\begin{equation}
q =\frac{1}{2}\Omega_M - \Omega_\Lambda \simeq -0.57 < 0~, 
\qquad (\Omega_\Lambda \simeq 0.7)~.
\end{equation}
Hence (\ref{o1}) and (\ref{o2}) provide evidence for a  
{\it current era acceleration of the Universe. }
At this stage it should be 
stressed that the recent observation of two more supernovae 
at $z > 1$~\cite{recentsn} 
supports the geometrical interpretation on the existence
of a dark energy component of the Universe, and 
argues rather against the r\^ole 
of nuclear (evolution) or intergalactic dust effects. 

The recent data of WMAP satellite  
lead to a new determination of 
$\Omega_{\rm total} =
1.02 \pm 0.02 $, where $\Omega _{\rm total} = \rho_{\rm total}/\rho_c$, due 
to high precision measurements 
of secondary (two more) acoustic peaks 
as compared with previous CMB measurements. Essentially the value of $\Omega$ 
is determined by the position of the first acoustic peak in a 
Gaussian model, whose reliability increases significantly 
by the 
discovery of secondary peaks and their excellent fit with the Gaussian 
model~\cite{spergel}.

\begin{flushleft}
{2. CONCLUSIONS}
\end{flushleft}


This paper provides a review of the variants of dark matter which are thought to be fundamental components of the universe and their role in origin and evolution of structures. 
It moreover gives some new original results concerning improvements to the spherical collapse model. In particular, how the spherical collapse model is modified when we take into account
dynamical friction and tidal torques. 
Studies of several decades have shown that, if we have a right knowledge of the law of gravity, dark matter is a fundamental component of our universe. While models based upon Hot Dark Matter (e.g., neutrinos) gives a reasonable description of structures on large scales
models based upon Cold Dark Matter (e.g., axions) are more successful 
in describing small and intermediate scales.
A fundamental ingredient in the recipe of structure formation is 
inflation which provides a spectrum of adiabatic Gaussian
perturbations which can be well described by a power-law spectrum, tilted from the Harrison--Zel'dovich spectrum, normally tilted so as to provide extra large scale power. The magnitude of the tilt may be modest or pronounced.
The details of structure formation are very sensitive to the matter content of
the universe. It appears that if cold dark matter is the main constituent of
the universe, present observations require that the initial perturbations be
adiabatic --- isocurvature perturbations generate excessively large cmb
anisotropies for the same final density perturbation. Adiabatic perturbations
are exactly what inflation provides. In CDM models, the only remaining
alternative would appear to be texture seeded models, which have been placed
in jeopardy by a combination of microwave anisotropy and velocity data, though
the death blow apparently remains to be struck.
%
%
The survey was completed by examining variants on the CDM model which may be
better suited to explaining the observational data. The standard technique is
to utilize additional matter (be it a component of hot dark matter or of a
cosmological constant) to remove short-scale power from the CDM spectrum. Hot
dark matter does this by free-streaming, a cosmological constant by delaying
matter-radiation equality. Because this power can be removed over a much
shorter range of scales than with tilt, it enables an explanation of the
observed deficit of short-scale power relative to intermediate scale power in
the spectrum.

MDM (Mixed dark Model) adds yet another new parameter, roughly speaking an ability
to remove short-scale power from the spectrum while leaving large scales
untouched, and may be necessary should all present observations stand up. It
appears likely that MDM will however need an initial spectrum close to $n=1$
with no gravitational waves if it is to succeed. 
Some studies (\cite{pee3};
Efstathiou et al. 1990a; \cite{tur}) has shown that  
lowering the matter density under
the critical value ($ \Omega_m < 1$) and adding a
cosmological constant in order to retain 
a flat Universe ($ \Omega _m + \Omega _\Lambda = 1$), gives good results in the case of $\Omega _m=0.3$. 
Moreover new observational evidences (see \cite{cora}) indicates that we are living in a $\Lambda \neq 0$ universe.

\begin{flushleft}
{AKNOWLEDGMENTS}
\end{flushleft}

The author thanks the referee M. V. Sazhin for his helpful comments.

~\\
~\\
~\\
~\\
~\\
~\\
~\\
~\\
~\\
~\\
\begin{figure}
\centerline{\hbox{
\psfig{file=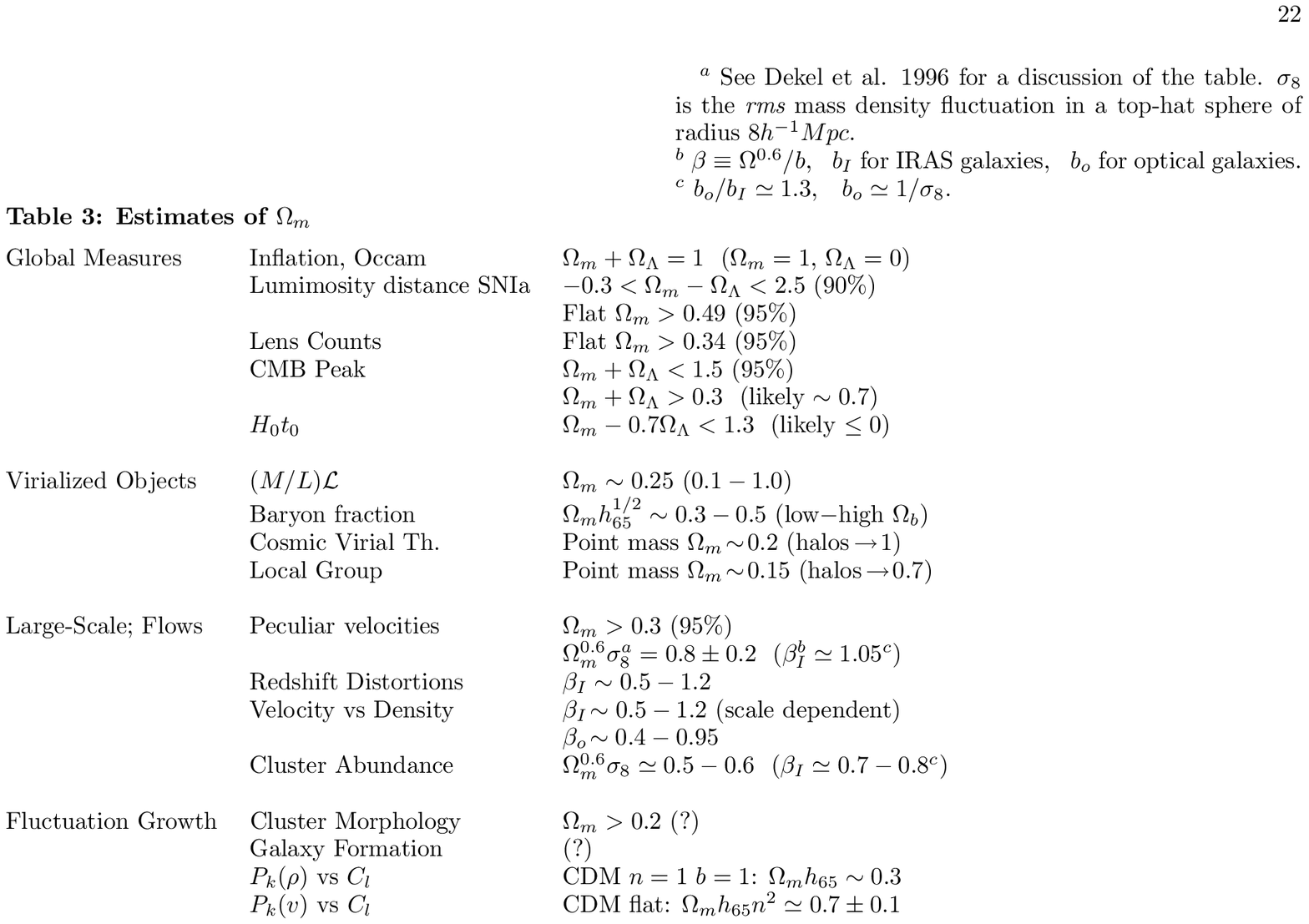,width=11.3cm}  
}}
\end{figure}

\end{document}